%% file: paper.tex
\documentclass[10pt,journal,compsoc]{IEEEtran}

\ifCLASSOPTIONcompsoc
  
  \usepackage[nocompress]{cite}
\else
  
  \usepackage{cite}
\fi

\usepackage{mathtools}

\usepackage{subcaption}
\usepackage{overpic}
\usepackage{pgfplots}

\usepackage{filecontents}
\usepackage{tikz,graphicx}
\usetikzlibrary{positioning}

\begin{document}

\title{A Voxel-based Rendering Pipeline for\\ Large 3D Line Sets}

\author{Mathias~Kanzler,
        Marc Rautenhaus,
        and~R\"udiger Westermann
\thanks{ M. Kanzler, M. Rautenhaus and R. Westermann
 are with the Computer Graphics \& Visualization Group, Technische Universit\"at M\"unchen, Garching, Germany.\protect\\
E-mail: \{mathias.kanzler, marc.rautenhaus, westermann\}@tum.de}
}

\IEEEtitleabstractindextext{
\begin{abstract}
We present a voxel-based rendering pipeline for large 3D line sets that employs GPU ray-casting to achieve scalable rendering including transparency and global illumination effects that cannot be achieved with GPU rasterization. Even for opaque lines we demonstrate superior rendering performance compared to GPU rasterization of lines, and when transparency is used we can interactively render large amounts of lines that are infeasible to be rendered via rasterization. To achieve this, we propose a direction-preserving encoding of lines into a regular voxel grid, along with the quantization of directions using face-to-face connectivity in this grid. On the regular grid structure, parallel GPU ray-casting is used to determine visible fragments in correct visibility order. To enable interactive rendering of global illumination effects like low-frequency shadows and ambient occlusions, illumination simulation is performed during ray-casting on a level-of-detail (LoD) line representation that considers the number of lines and their lengths per voxel. In this way we can render effects which are very difficult to render via GPU rasterization. A detailed performance and quality evaluation compares our approach to rasterization-based rendering of lines.

\end{abstract}

\begin{IEEEkeywords}

Ray-casting, large 3D line sets, transparency, global illumination
\end{IEEEkeywords}}

\maketitle

\IEEEdisplaynontitleabstractindextext

\IEEEpeerreviewmaketitle

\IEEEraisesectionheading{\section{Introduction}\label{sec:introduction}}

 \input{sections/introduction.tex}

\input{sections/voxelization.tex}

 \input{sections/results.tex}

\input{sections/conclusion.tex}

\input{sections/colorpage.tex}

\bibliographystyle{IEEEtran}

\bibliography{LineRaycast}

\vspace*{-1cm}
\begin{IEEEbiography}[{\includegraphics[width=1in,height=1.25in,clip,keepaspectratio]
{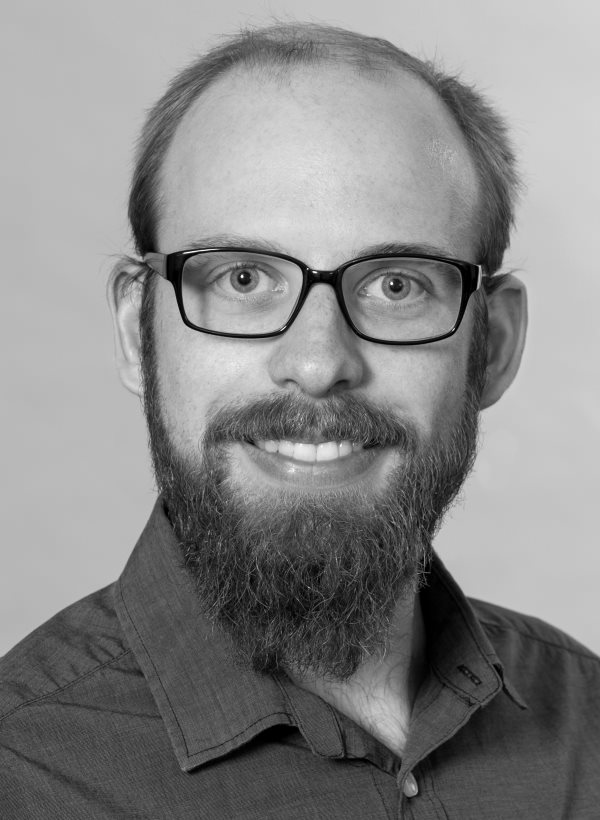}}]{Mathias Kanzler}
is a PhD candidate in the Computer Graphics and Visualization Group at the Technical University of Munich 
(TUM). He received the M.Sc. in computer science from TUM in 2015. His research
interests include visualization and real-time rendering.

\end{IEEEbiography}

\vspace*{-1cm}
\begin{IEEEbiography}[{\includegraphics[width=1in,height=1.25in,clip,keepaspectratio]
{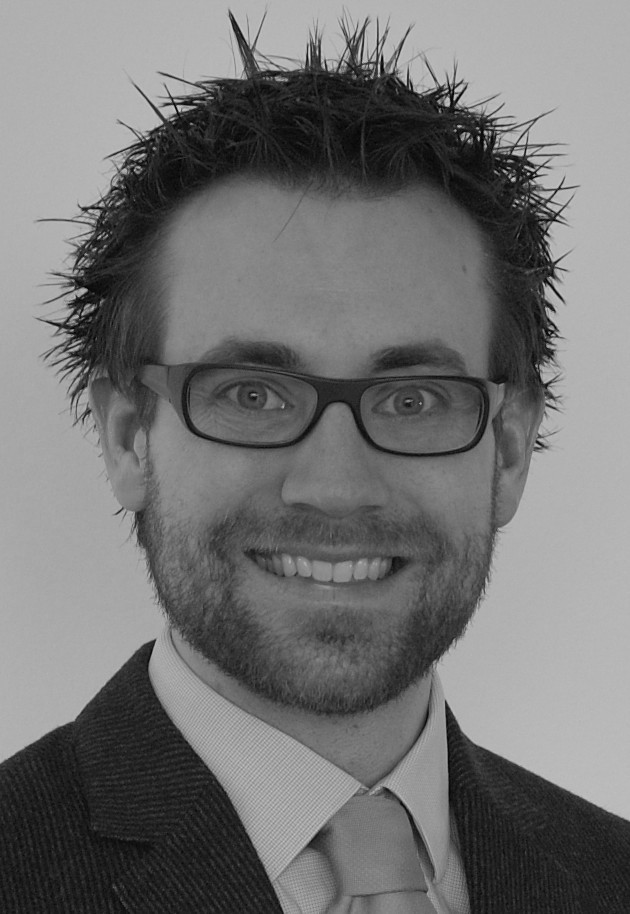}}]{Marc Rautenhaus}
is a postdoctoral researcher in the 
Computer Graphics and Visualization Group at the TUM. He received the M.Sc. in atmospheric science from the University of 
British Columbia, Vancouver, in 2007, and the Ph.D. in computer science from 
TUM in 2015. Prior to joining TUM, Marc worked as a research associate at the
German Aerospace Center's Institute for Atmospheric Physics. His research 
interests focus on the intersection of visualization and meteorology.
\end{IEEEbiography}

\vspace*{-1cm}
\begin{IEEEbiography}[{\includegraphics[width=1in,height=1.25in,clip,keepaspectratio]
{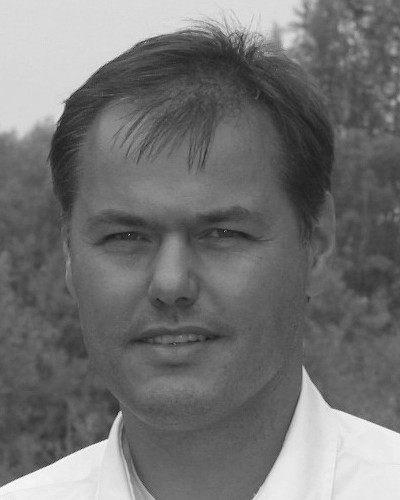}}]{R\"udiger Westermann}
studied computer science at the Technical 
University Darmstadt and received the Ph.D. in computer science from the 
University of Dortmund, both in Germany. In 2002, he was appointed the chair of 
Computer Graphics and Visualization at TUM. His research interests include 
scalable data visualization and simulation algorithms, GPU computing, real-time 
rendering of large data, and uncertainty visualization.
\end{IEEEbiography}

\end{document}

%% file: sections/introduction.tex
3D line sets can be rendered efficiently via the rasterization-based rendering pipeline on GPUs, for instance, by constructing tubes around each line on-the-fly in a geometry shader. For the rendering of very large 3D line sets, however, GPU rasterization introduces the following limitations: a) When using transparency, which requires the rendered fragments to be blended in correct visibility order, all fragments need to be stored in a per-pixel linked list~\cite{Thibieroz2010} on the GPU and sorted with respect to the current camera position. This approach works efficient for moderate sized line sets, yet for large sets the sorting operation significantly slows down performance, and in some cases the lists can even exceed the GPU's memory (see Fig.~\ref{fig:out-of-memory}). b) Operations requiring spatial adjacency queries cannot be embedded efficiently into rasterization-based rendering. For instance, shadow simulation or ambient occlusion calculations to enhance the spatial perception of rendered lines. Even though hard shadows of point lights can be simulated via shadow mapping, this requires a second rendering pass using the full set of geometry, and the resulting high-frequency illumination variations are rather disturbing from a perceptual point of view (see Fig.~\ref{fig:hardshadows}).

An option to overcome these limitations is a rendering technique that does not build upon the order-independent projection of primitives, but is ray-guided and can efficiently traverse the line set along an arbitrary direction. For the rendering of polygon models, GPU voxel ray-casting has been established as a powerful alternative to rasterization. Voxel ray-casting employs a voxel-based object representation in combination with a regular sampling grid that can be traversed efficiently on the GPU~\cite{Gigavoxel2009,Laine2010}. This approach is particular useful because it can generate the sample points along a ray in an order-dependent way, and provides the ability to perform adaptive LoD selection. Furthermore, the regular sampling grid gives rise to efficient search operations, which have been used to simulate global illumination effects via ray-guided visibility computations~\cite{Crassin:2011:III:1944745.1944787,Thiedemann:2011:VGI:1944745.1944763}.

Mimicking voxel ray-casting for line primitives, however, is challenging, since no voxel representation of lines is known that can effectively encode spatial occupancy \emph{and} direction. A solid voxelization of line primitives---or tubes generated from them---into a regular sampling grid~\cite{Cohen-Or:1997:LVC:616049.618482} is not feasible due to the following reasons: Firstly, for the line sets we consider, too many voxels must be generated to differentiate between individual lines. Secondly, upon voxelization, changes of rendering parameters like line width and line removal require exhaustive processing. Thirdly, already at moderate zoom factors the lines will appear blocky. Finally, since voxels do not encode any directional information, direction-guided shading and filtering as well as direction-preserving LoD construction becomes infeasible.

\subsection{Contribution}

In this work we propose an alternative rendering technique for large 3D line sets on the GPU, which builds upon the concept of voxel ray-casting to overcome some of the limitations of rasterization-based line rendering. We present a new voxel model for 3D lines and demonstrate its use for GPU line rendering including transparency and ambient occlusion effects. We further show that the voxel model can be used in combination with rasterization-based rendering, by performing the simulation of volumetric effects on this model and letting the rendered fragments access the computed results.

Our specific contributions are:
\begin{itemize}
\item A novel voxel-based representation of lines, consisting of a macro voxel grid, and a per-voxel quantization structure using voxel face-to-face connectivity.
\item A LoD line representation that considers per voxel the number of lines, their orientations, and lengths to estimate an average line density and direction at ever coarser resolutions.
\item An implementation of GPU ray-casting on the voxel-based representation, including the computation of line, respectively tube intersections on the voxel level, to efficiently simulate transparency as well as local and global illumination effects. 
\end{itemize}

We see our rendering approach as an alternative to existing rendering techniques for 3D line sets when there are many lines introducing massive overdraw, and when transparency or global illumination effects are used to enhance the visual perception (see Fig.~\ref{fig:colorpage-light}). In this case, our method can significantly accelerate the rendering process, and it can even be used when the memory requirements of rasterization-based approaches exceed what is available on current GPUs. The possibility to construct a LoD line representation enables trading flexibly between quality and speed, and efficiently performing search operations required for simulating advanced illumination effects.

The remainder of this paper is organized as follows: We first introduce the particular voxel-model underlying our approach. Next, we show how this model is ray-traced, and how transparency and global illumination effects are integrated. We further demonstrate the embedding of line ray-casting in rasterization-based line rendering approaches. Then, we perform a thorough evaluation of our approach regarding quality, speed and memory requirement, and we demonstrate its benefits and limitations with data sets from different applications. We conclude our work with some ideas on future extensions and applications.

\section{Related work}

Our approach is related to established techniques in visualization and rendering, namely line-based rendering of flow fields and voxel-based ray-tracing.

\subsection*{Line-based rendering of flow fields}

For the visualization of integral curves in vector fields, numerous techniques have been proposed to improve their visual perception~\cite{CGF:CGF1650}. Today, 3D line sets are usually visualized via the rasterization-based rendering pipeline on GPUs, either as illuminated line primitives~\cite{Zockler:1996:IVF:244979.245023,1532772}, or as tubes which are constructed around each line on-the-fly in the GPU's shader units.

For dense sets of opaque lines, illustrative rendering ~\cite{5290742} has been proposed to keep separate lines perceptually visible. Other techniques abstract from single primitives and show flow features via line density projections\cite{Park:2006:SDF:2384796.2384821,Kuhn:2013:EUROVIS}. Automatic view-dependent selection of transparency has been introduced by G\"unther et al.~\cite{Gunther:2013:OOL:2461912.2461930}, to selectively fade out lines in those regions where they occlude more important ones. A user study on perceptual limits of transparency-based line rendering for flow visualization has been conducted by Mishchenko and Crawfis~\cite{CGF:CGF12268}. They also suggest a number of specific usages of transparency to effectively avoid visual clutter and high levels of occlusions.

When transparent lines are rendered, generated line fragments need to be blended in correct visibility order. On the GPU, this can be realized by using either depth peeling~\cite{Everitt2001InteractiveOT} or per-pixel fragment lists~\cite{Thibieroz2010,Yang:2010:RCL:2383616.2383624}. Depth peeling does not require storing and sorting fragments, yet it requires rendering the data set as many times as the maximum depth complexity of the scene, i.e., the maximum number of fragments falling into the same pixel. For the data sets we consider, where along the majority of view rays the depth complexity is in the order of many hundred or even thousand, the resulting increase in rendering times is not acceptable. Per-pixel fragment lists, on the other hand, require only one single rendering pass, yet they require storing all fragments and can quickly run out of memory on current GPUs. This also holds when depth peeling is applied to fragments bucketed by depth~\cite{Liu:2009:EDP:1572769.1572779}, even though the number of rendering passes can be reduced. As an alternative to the exact simulation of transparency, stochastic transparency~\cite{5601714} uses sub-pixel masks to stochastically sample the fragments' transparency contributions. Stochastic transparency requires only a rather small and fix number of rendering passes, yet it transforms directional structures into noise. This is especially undesirable in our scenarios, where only lines are rendered and even in transparent regions their directional structure should be preserved.

The technique most closely related to ours is the one by Schussman and Ma~\cite{Schussman:2004:AVR:1032664.1034437}. They voxelize streamlines into a regular grid, and compute a spherical harmonics representation of the lighting distribution caused by the line segments in every voxel. Voxel-based rendering with sub-pixel illumination can then be performed, yet single line primitives cannot be determined any more, for instance, to render the initial lines or compute exact occlusions.

\subsection*{Voxel-based ray-tracing}

Advances in hardware and software technology have shown the potential of ray-tracing as an alternative to rasterization, especially for high-resolution models with many inherent occlusions. Developments in this field include advanced space partitioning and traversal schemes~\cite{Wald:2006:RTAS,Woop:2006:BKD,Wald:2017:OCR:3024362.3025410}, and optimized GPU implementations
\cite{Aila2009hpg,Laine:2010:ESV:1730804.1730814,Parker:2010:OGP:1778765.1778803}, to name just a few.
All these approaches can be classified as ``conventional ray-tracing approaches'', since they operate on
the polygon object representation and perform classical ray-polygon intersection tests. Recently, Wald et al.~\cite{conf/scivis/WaldKJUPP15} proposed the use of ray-tracing for particle rendering, by using a tree-based search structure for particle locations to efficiently find those particles a ray has to be intersected with. The integration of global illumination effects like ambient occlusion into particle visualizations has been demonstrated by Wald et al.~\cite{conf/scivis/WaldKJUPP15} and Staib et al~\cite{CGF:CGF12627}.

Voxel models have been first introduced 1993 by Kaufman in the seminal paper \cite{Kaufman1993}. Compared to polygonal representations, they provide an interesting set of advantages like easier level of detail computation and combined storage of surface and geometry information. A detailed investigation of algorithms for line voxelization has been provided by Cohen-Or and Kaufman~\cite{Cohen-Or:1997:LVC:616049.618482}. Interestingly, a voxel model has also been used in the very first published work on iso-surface visualization: \emph{Cuberilles} \cite{Herman1979}. The Cuberille method---or opaque cubes---works by computing the set of grid cells that contain a selected iso-surface and rendering those as small cubes.

In the last years, there has been a lot of research into large octrees to render high-resolution voxel models~\cite{Gigavoxel2009,Laine2010,VMV12:47-54:2012}. All these approaches subdivide the model into a sparse volume, storing only small volume ``bricks'' along the initial model surface, and use a compute-based octree traversal to render the contained surface. The potential of voxel-based rendering approaches to efficiently simulate global illumination effects has been demonstrated, for instance, in ~\cite{CNSGE11b,Thiedemann:2011:VGI:1944745.1944763}. The survey by Joenssen et al~\cite{CGF:CGF12252} provides a general overview of approaches for simulating global illumination effects on volumetric data sets.

%% file: sections/voxelization.tex
\section{Voxel-based curve discretization}

\begin{figure}[h]
\begin{overpic}[width=.49\textwidth]{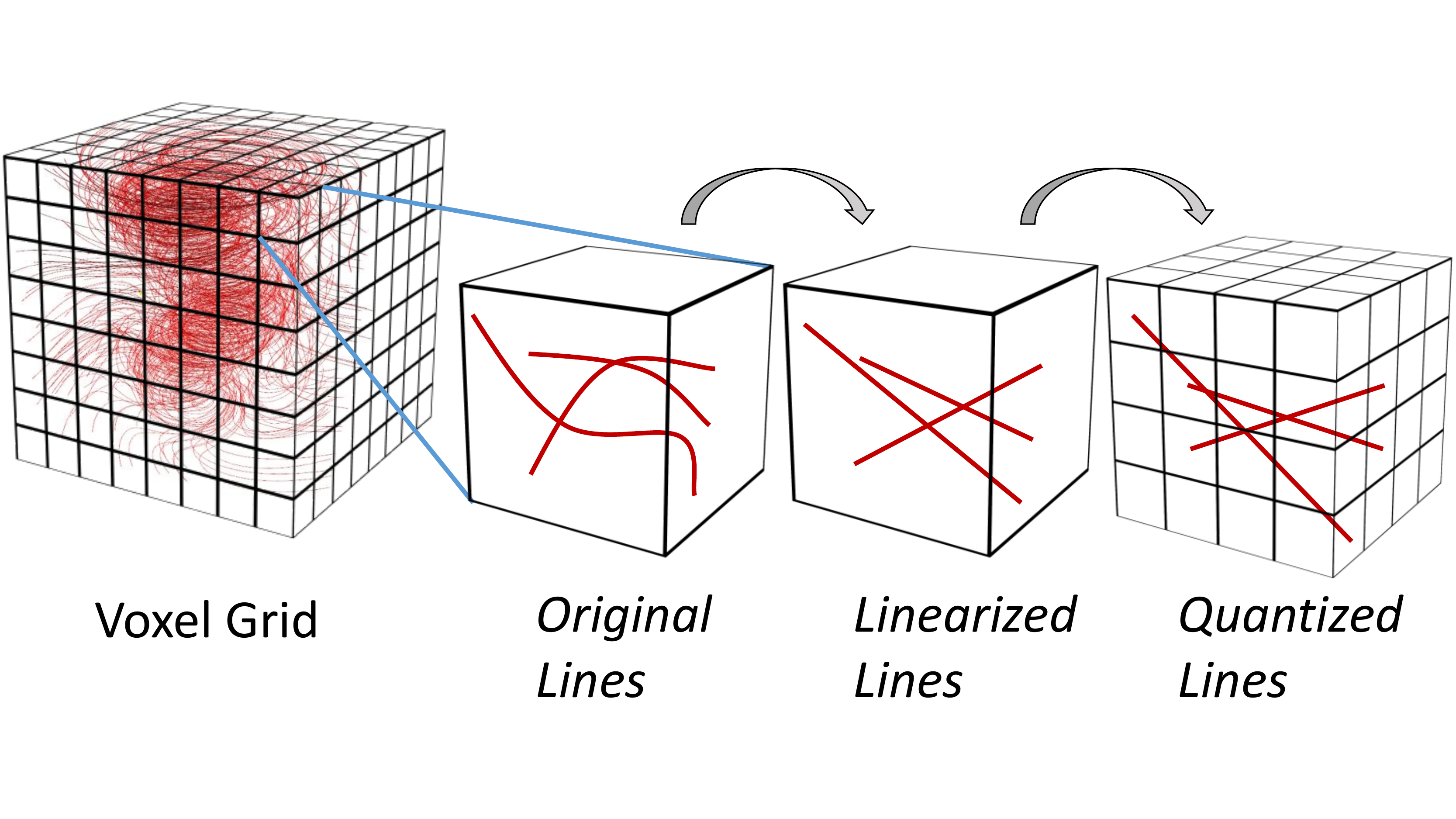}
	\put(0,47){\textbf{\textcolor{black}{(a)}}}
	\put(35,47){\textbf{\textcolor{black}{(b)}}}
	\put(60,47){\textbf{\textcolor{black}{(c)}}}
	\put(82,47){\textbf{\textcolor{black}{(d)}}}
\end{overpic}
\vspace*{-10mm}
\caption{Two-level grid structure. First level: Voxel grid, curves are clipped to the voxel faces (b), and per-voxel linear line segments are generated (c). Second level: Line vertices are quantized based on a uniform subdivision of voxel faces (d).}
\label{fig:two-level-structure}
\end{figure}
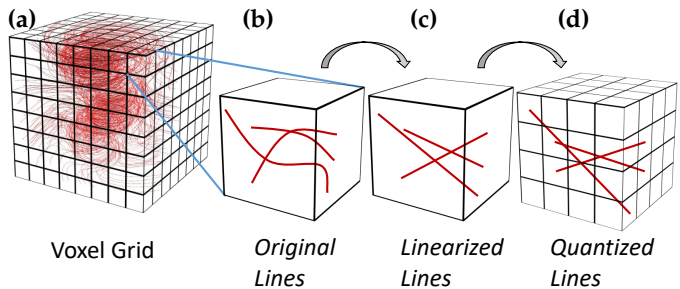

The input of our method consists of a set of curves, e.g. streamlines, where each curve is approximated by a connected line set. Each line set is represented by a sequence of vertices $v_i$ and corresponding attributes $a_i$. The attributes can be specified optionally, for instance, to enable the use of a transfer function to interactively change the curves' colors or transparencies. In the following we assume that the curve discretization is performed in a pre-process on the CPU, and the generated data structure is then uploaded to the GPU where rendering is performed. This pre-process can also be performed on the GPU, enabling instant update operations when curves are removed or new curves are added. This, however, requires some modifications, the discussion of which we delay until the end of the current section.
We propose a two-level grid structure for discretization of the initial curves, as illustrated in Fig.~\ref{fig:two-level-structure}.

The first level is given by a voxel grid, in which each curve is approximated by a set of per-voxel linear line segments. A regular subdivision of each voxel face yields the second level, where the endpoints of the per-voxel line segments are quantized to the center points of sub-faces.

The use of a two-level structure allows controlling the approximation quality of the voxelization and storing the line segments per voxel in a compact form. The resolution of the voxel grid controls the size of the geometric features that can get lost when approximating these features by linear segments in every voxel. With increasing grid resolution better approximation quality is achieved, yet at the cost of increasing memory requirement. The end points of each per-voxel line segment can either be stored exactly using floating point values, or they can be quantized to a set of points on the voxel faces. This allows for a compact encoding of the line segments per voxel, and it can be used to control how many lines passing through one voxel are collapsed to one single line.

\subsection{Curve voxelization}

Initially, the resolution $r_x \times r_y \times r_z$ of the 3D voxel grid into which the curves are voxelized is set. We use cube-shaped voxels of side length 1 and set the resolution so that the aspect ratio of the bounding box of the initial curves is maintained. The vertex coordinates $v_i$ are then transformed to local object coordinates in the range from $(0,0,0)$ to $(r_x, r_y, r_z)$.

For each curve and starting with the first vertex, every pair of vertices $v_i$ and $v_{i+1}$ is processed consecutively. A line through $v_i$ and $v_{i+1}$ is clipped against the voxel boundaries, via line-face intersection tests in the order of their occurrence from $v_i$ to $v_{i+1}$. If $v_i$ and $v_{i+1}$ are located in the same voxel, no new intersection point is generated. This gives a sequence of voxel-face intersections, and every pair of consecutive intersections represents a line that enters into a voxel and exits that voxel. In general the first and last vertex of a line do not lie on a face, we hence omit the segments from the first vertex to the first face intersection and from the last face intersection to the last vertex.

\begin{figure}[h]
\includegraphics[width=.49\textwidth]{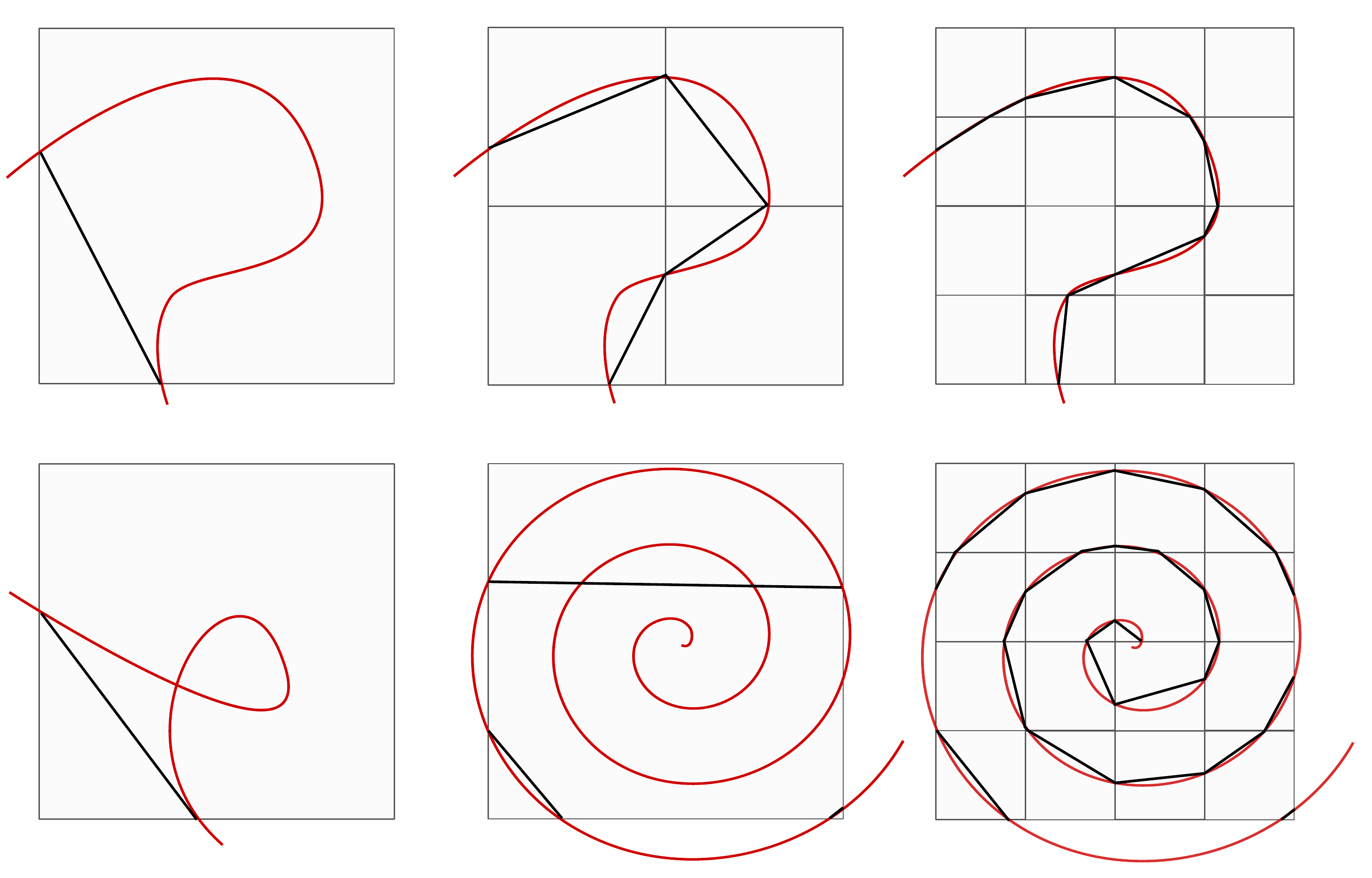}
\caption{Curve voxelization: Initial curve in red and piecewise linear approximation in black. Top: Decreasing approximation error with increasing voxel grid resolution. Bottom: Geometric details can be missed if the voxel grid resolution is too low.}
\label{fig:curve_discretization}
\end{figure}

Fig.~\ref{fig:curve_discretization} illustrates the voxelization process, demonstrating increasing approximation quality with increasing grid resolution, as well as limitations of the piecewise linear curve approximation when the grid resolution is too low. 
The maximal deviation between the initial curve and the generated line segments is bound by the length of a voxel's diagonal.

For every voxel in the voxel grid, a linked list is used to store the lines intersecting that voxel. Line attributes, interpolated to the line-face intersection points, and IDs that identify to which curve a line belongs can be stored in addition.

\subsection{Line quantization}

\begin{figure}[h]
\includegraphics[width=.49\textwidth]{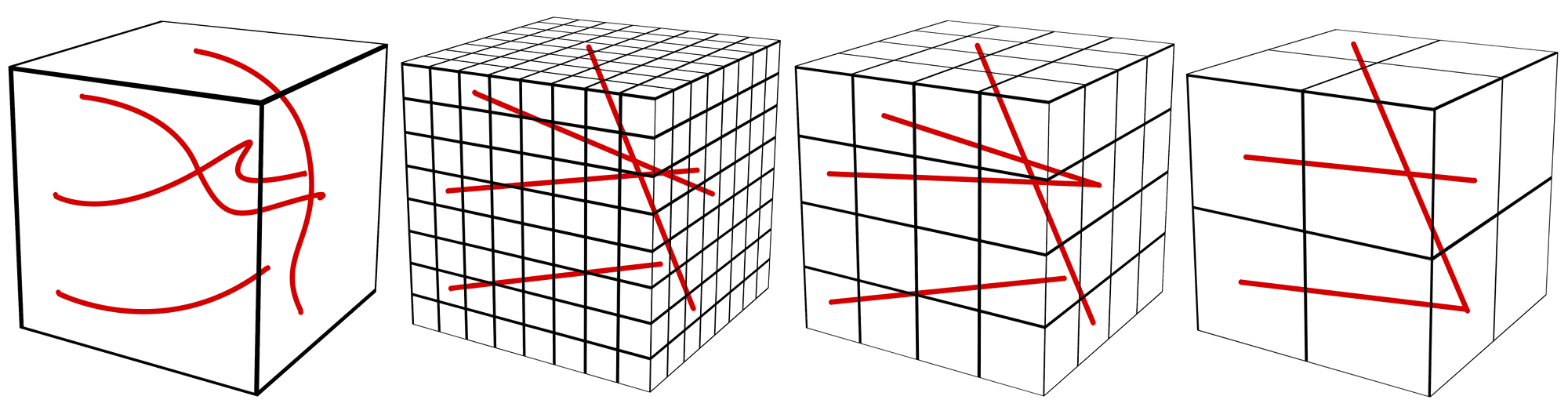}
\caption{Line quantization. From left to right: Original curves, per-voxel linear approximation using vertex quantization to $8^2, 4^2,$ and $2^2$ sub-faces (bins) per voxel face, respectively. If the number of bins is too low, lines can fall onto each other. }
\label{fig:voxelization}
\end{figure}

For quantizing the coordinates of the curve-voxel intersection points, we use a small number of bins per face: As illustrated in Fig~\ref{fig:two-level-structure}d, every square voxel face is subdivided regularly into $N \times N$ smaller squares. The coordinates of the intersection points are then quantized to the centers of these squares. Thus, every vertex can be encoded via a $2N$ bit pattern (indicating the sub-square in a voxel face) and a 3 bit face ID (indicating on which face the vertex is located). The bits are packed into a single word that is just large enough to store them. Since voxels have unit size and each voxel face is subdivided equally, the vertex location in object coordinates (quantized to the per-face square centers) can be computed from a bit pattern stored per-vertex. Fig.~\ref{fig:voxelization} illustrates the quantization process and the effects of different values of $N$ on the curve approximations.

The quantization process introduces an additional approximation error that depends on the selected subdivision of voxel faces. This error does not cause any geometric details to be lost, yet it slightly jitters the vertex locations and, thus, affects a line's orientation. In particular, different lines might be mapped onto the same quantized line because their vertices are quantized to the same locations.

\begin{figure}[h]
\includegraphics[width=0.24\textwidth]{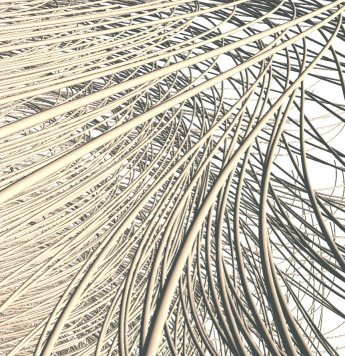}\hfill
\includegraphics[width=0.24\textwidth]{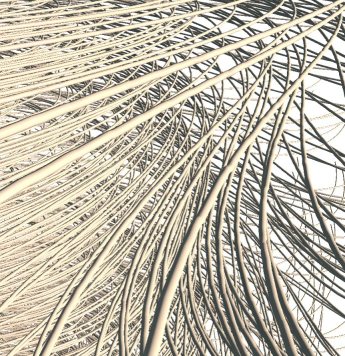}
\caption{GPU rasterization of initial lines (left) vs. rasterization of voxelized lines using $256^3$ voxel grid and curve-face intersections quantized to $16^2$ bins (right).}
\label{fig:rast-vs-rast}
\end{figure}

Fig.~\ref{fig:rast-vs-rast} shows a direct comparison between the original curves and the voxelized curves using a voxel grid resolution and per-face subdivision at which the discretization errors can only just be perceived. In both cases, the lines are rendered via GPU rasterization; by starting at the first vertex and then either by traversing the original set of vertices and constructing tubes around each line on-the-fly in a geometry shader, or by performing the same construction on the discretized line set. Especially the straight curves in the foreground show some subtle bumpiness that is caused by the quantization of line vertices.

\begin{figure*}[h]
\begin{overpic}[width=0.24\textwidth]{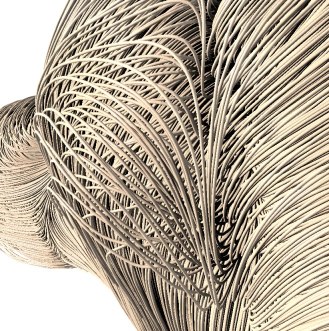}
	\put(4,2){ {\textcolor{black}{a)}}}
\end{overpic}\hfill
\begin{overpic}[width=0.24\textwidth]{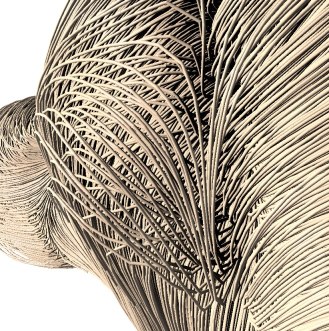}
	\put(4,2){ {\textcolor{black}{b)}}}
\end{overpic}\hfill
\begin{overpic}[width=0.24\textwidth]{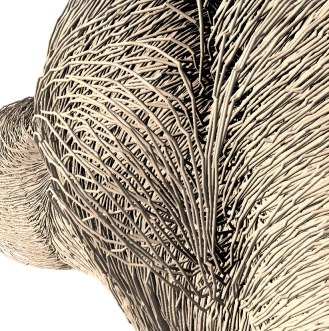}
	\put(4,2){ {\textcolor{black}{c)}}}
\end{overpic}\hfill
\begin{overpic}[width=0.24\textwidth]{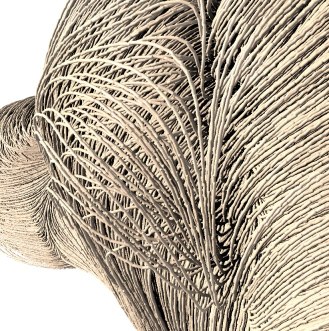}
	\put(4,2){ {\textcolor{black}{d)}}}
\end{overpic}
\caption{Streamlines in the $256^3$ Aneurysm II dataset. (a) Initial lines. (b) Line encoding in $64^3$ voxel grid with exact curve-voxel intersections. (c) Line encoding in $64^3$ voxel grid with curve-voxel intersections quantized to $8^2$ bins. (d) Line encoding in $256^3$ voxel grid with curve-voxel intersections quantized to $8^2$ bins.}
\label{fig:NearQuantization}
\end{figure*}

To demonstrate how, in general, the voxel grid resolution and the voxel face subdivision affect the final reconstruction quality, Fig.~\ref{fig:NearQuantization} shows some extreme cases which also reveal the interplay between both resolutions. A detailed analysis of the dependencies between resolution, quality and memory consumption for different resolutions is given in Sec.~\ref{sec:results}. While a low resolution of the voxel grid affects in particular the per-voxel approximation error, a low degree of face subdivision can introduce additional $C^1$-discontinuities at voxel transitions, i.e., by jittering a vertex that is shared by two lines with the same orientation. Especially high frequent variations that occur when a short line is jittered are perceptually noticeable, as indicated by the last example in Fig.~\ref{fig:NearQuantization}.

\subsection{GPU implementation}

The per-voxel linked list representation used to store the lines is neither memory efficient, as each line stores a pointer to the following element, nor can subsequent lines be accessed quickly, as they are not stored consecutively in memory and, thus, cached reads are prohibited. Therefore, before uploading the data to the GPU, the lines are reordered in memory so that all lines passing through the same voxel are stored in consecutive elements of a linear array. The final data structure stores for each voxel a header, which stores the number of lines for that voxel, and a pointer to the array element storing the first line passing through it.

Constructing the voxel model on the GPU consists of three stages: Firstly, in parallel for every initial curve we compute the lines per voxel and write them into a linear append buffer. Every segment is assigned the unique ID of the voxel it is contained in. Next, the buffer is sorted with respect to the voxel IDs, so that all lines falling into the same voxel are located consecutively in the buffer. An exclusive parallel prefix sum is then computed in-place over the buffer to count the number of line per voxel. Now, for every voxel the start index of its line set can be determined from the content of the buffer, and stored in a separate voxel buffer at the corresponding location. This buffer is used at render time to look up at which position in the global buffer the lines for a particular voxel are stored.

\subsection{LoD construction}
\label{sec:LoD}

One important property of classical voxel models for surfaces is that a LoD structure can be generated efficiently by simple averaging operations on the voxel values to aggregate information. For a line set that is voxelized as proposed in our work, such an averaging operation cannot be applied immediately because a) an averaging operator for lines first needs to be defined, and b) every voxel might store not only one but many lines. It is worth noting that regardless of how a) and b) are addressed, it is impossible, in general, to represent the lines in one voxel by one single average line so that continuity with the average lines in adjacent voxels is ensured.

\begin{figure}[h]
\includegraphics[width=0.12\textwidth]{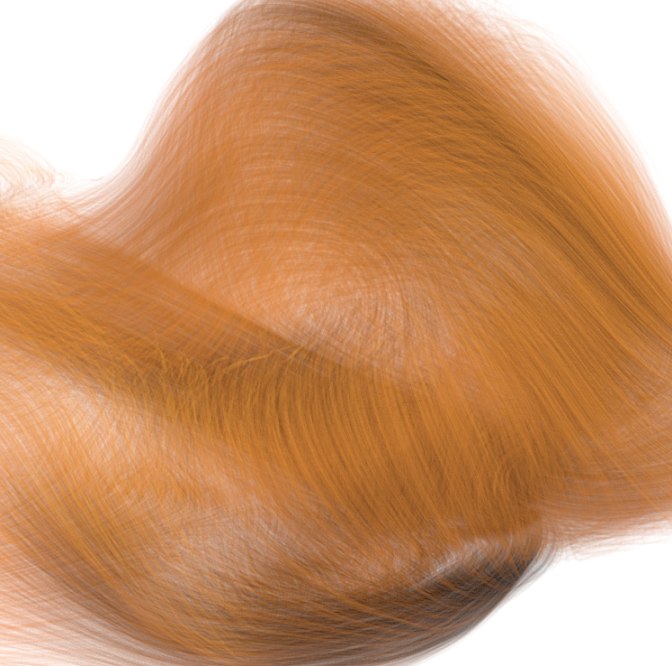}\hfill
\includegraphics[width=0.12\textwidth]{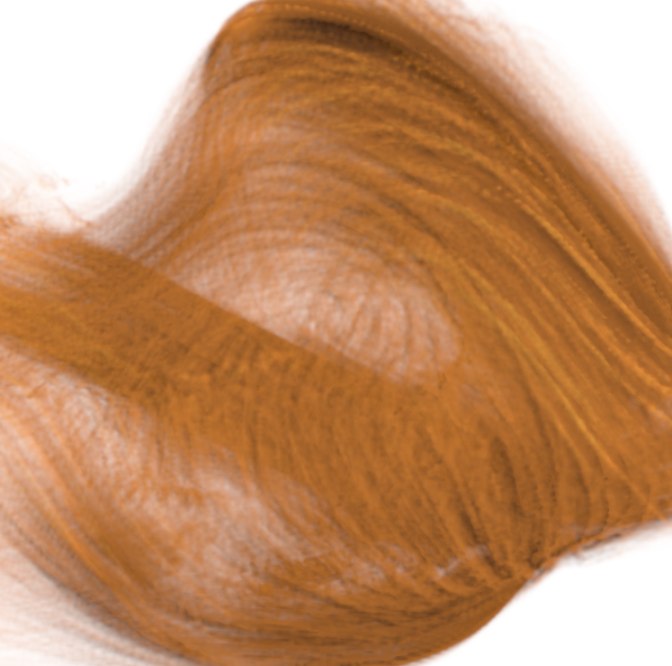}\hfill
\includegraphics[width=0.12\textwidth]{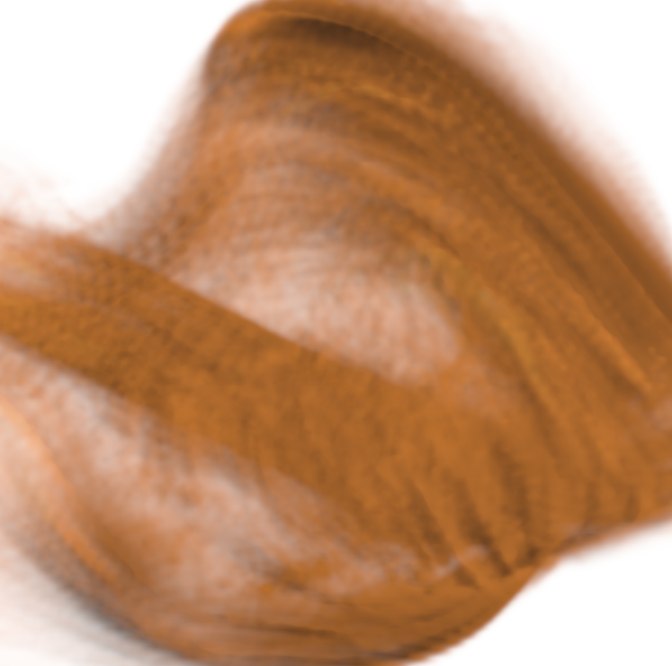}\hfill
\includegraphics[width=0.12\textwidth]{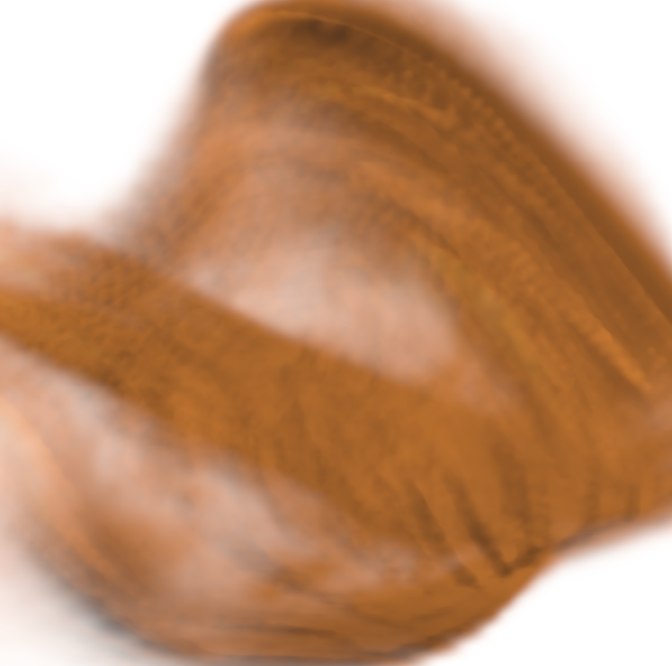}
\caption{Initial lines, line density values at the first, second, and third LoD.}
\label{fig:density}
\end{figure}

We address the problem of LoD construction for a voxelized line set as follows: Since we intend to use the LoD structure in particular to accelerate the simulation of global illumination effects, we derive a scalar indicator for the amount of light that is blocked by the lines in a single voxel. By using standard averaging operators, an octree LoD structure can then be generated in a straight forward way from the indicator field. For every voxel, we compute an average \emph{density} value $\rho$ from the lines passing through it, by taking into account the lines' lengths and opacities:
\begin{equation}
\rho=\sum_{l_i \in L}  length(l_i)\sigma_i,
\end{equation}
where $L$ is the set of lines in the voxel, and $l_i$ and $\sigma_i$ are the length and opacity of the i-th line in that voxel, respectively. The lines' opacities are either set to a constant values or assigned individually via a transfer function. Finally, an octree is build bottom-up by averaging the density values in $2^3$ voxels into one voxel at the next coarser level (see Fig.~\ref{fig:density}).

The density values can be interpreted as per-voxel opacities telling how much light is blocked by a voxel. Since the amount of blocking depends also on how the lines are oriented with respect to the incoming light, direction-dependent opacity values are favorable in principle~\cite{Gobbetti:2005:FVM:1073204.1073277}. Even though they can be generated in a straight forward way, by simulating the attenuation from a number of selected directions via ray-casting, we did not consider this option in the current work to avoid storing many values per voxel and, thus, increasing the memory consumption significantly.

\begin{figure}[h]
\includegraphics[width=0.15\textwidth]{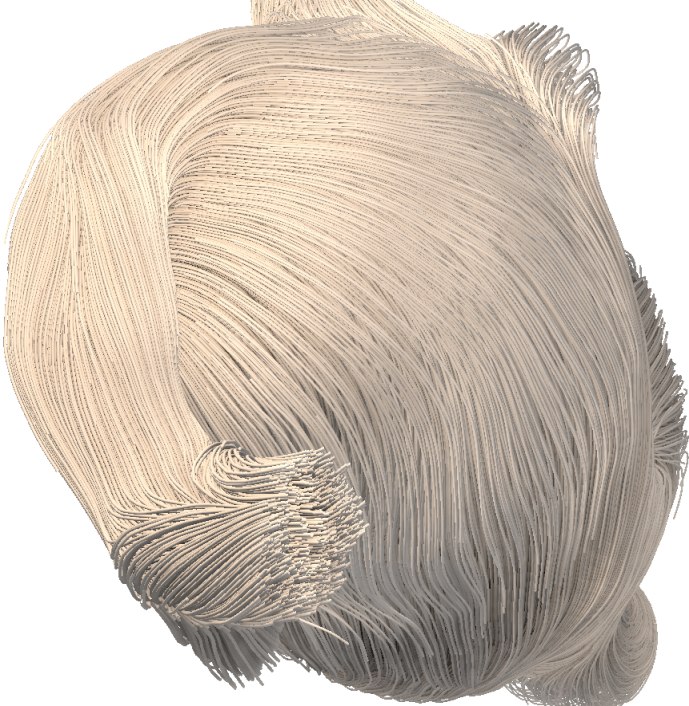}\hfill
\includegraphics[width=0.15\textwidth]{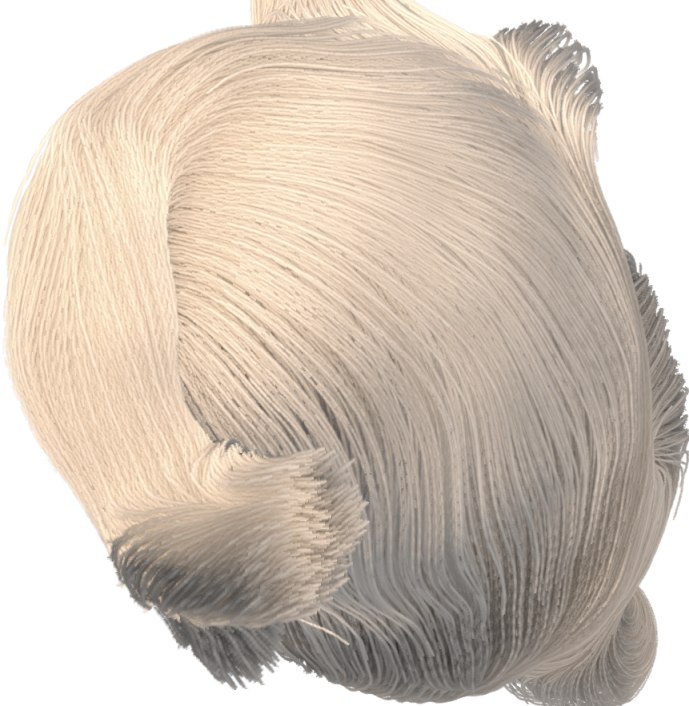}\hfill
\includegraphics[width=0.15\textwidth]{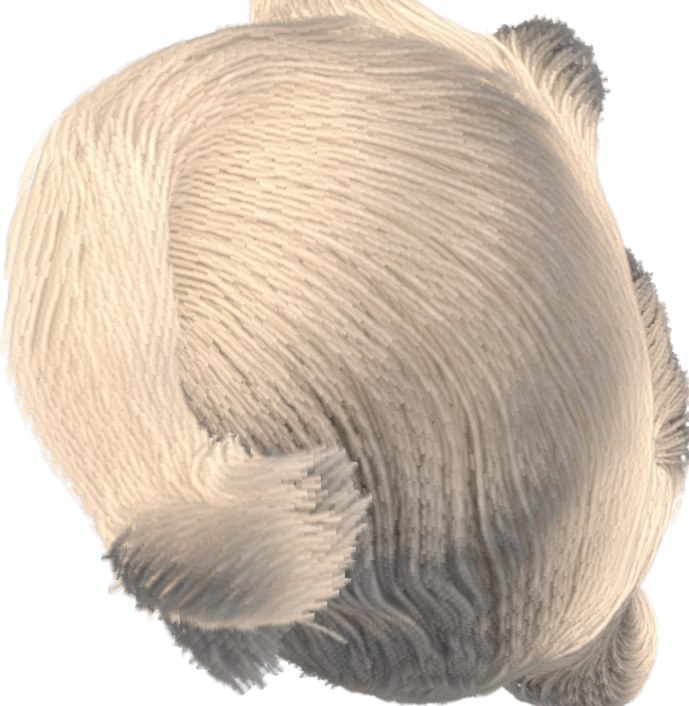}\hfill
\caption{Voxelized lines, first, and second LoD.}
\label{fig:lod}
\end{figure}

In addition to the per-voxel opacity values, one representative average line is computed for every voxel; by averaging separately the start points and end points of all lines in a voxel. This gives two average points which are snapped to the closest bin on the voxel boundaries, and from which the average line is computed. Care has to be taken regarding the orientation of lines, i.e., when two averaged lines have vastly opposite directions. Therefore, before considering a new pair of start and end point, we first test whether the corresponding line has an angle of more than $90°$ with the previous line, and we flip the line if this is the case. We also attempt to establish connectivity between representative lines in adjacent voxels if their endpoints are in the same voxel face, by snapping the endpoints to the quantization bin into which the average of both points is falling. In this way, we can often achieve continuous representative curves, even though it is clear that in general such a continuous fit is not possible. Fig.~\ref{fig:lod} illustrates a multi-resolution representation of a set of curves. Note in particular how well in certain regions the average curves even on the coarser resolution levels represent the initial curves.

The representative lines are stored in addition to the per-voxel line set, and a LoD representation can be computed by propagating these lines to ever coareser resolution levels. The average per-voxel opacity values at different resolution levels are stored in a set of 3D texture maps. By using these quantities we can efficiently realize a number of rendering accelerations and effects, which we will introduce in Sec.~\ref{sec:rendering}.

\section{Line Raycasting}

Once the voxel-based line representation has been constructed and the data is residing in GPU memory, volume ray-casting can be used to render the lines in front-to-back order. For every pixel a ray is cast through the voxel grid, thereby going from voxel face to voxel face using a digital differential analyzer algorithm. The voxel grid serves as a search structure to efficiently determine those lines that need to be tested for an intersection with the ray.

Whenever a voxel is hit, the voxel header is read to determine how many lines are stored in that voxel, and if a voxel is empty, it is skipped. Otherwise, the lines are read consecutively and intersected with the ray. Here it is assumed that the lines are in fact tubes with a user-defined radius, so that the intersection test becomes a ray-tube intersection test. This test yields an entry and exit point, from which the distance the ray travels inside the tube can be computed and used, for instance, to simulate attenuation effects. If more than one intersection with tubes are determined, the intersections are first computed and then sorted in place in the correct visibility order.

\subsection{Ray-tube intersections}
\begin{figure}[h]
\centering
\includegraphics[width=.23\textwidth]{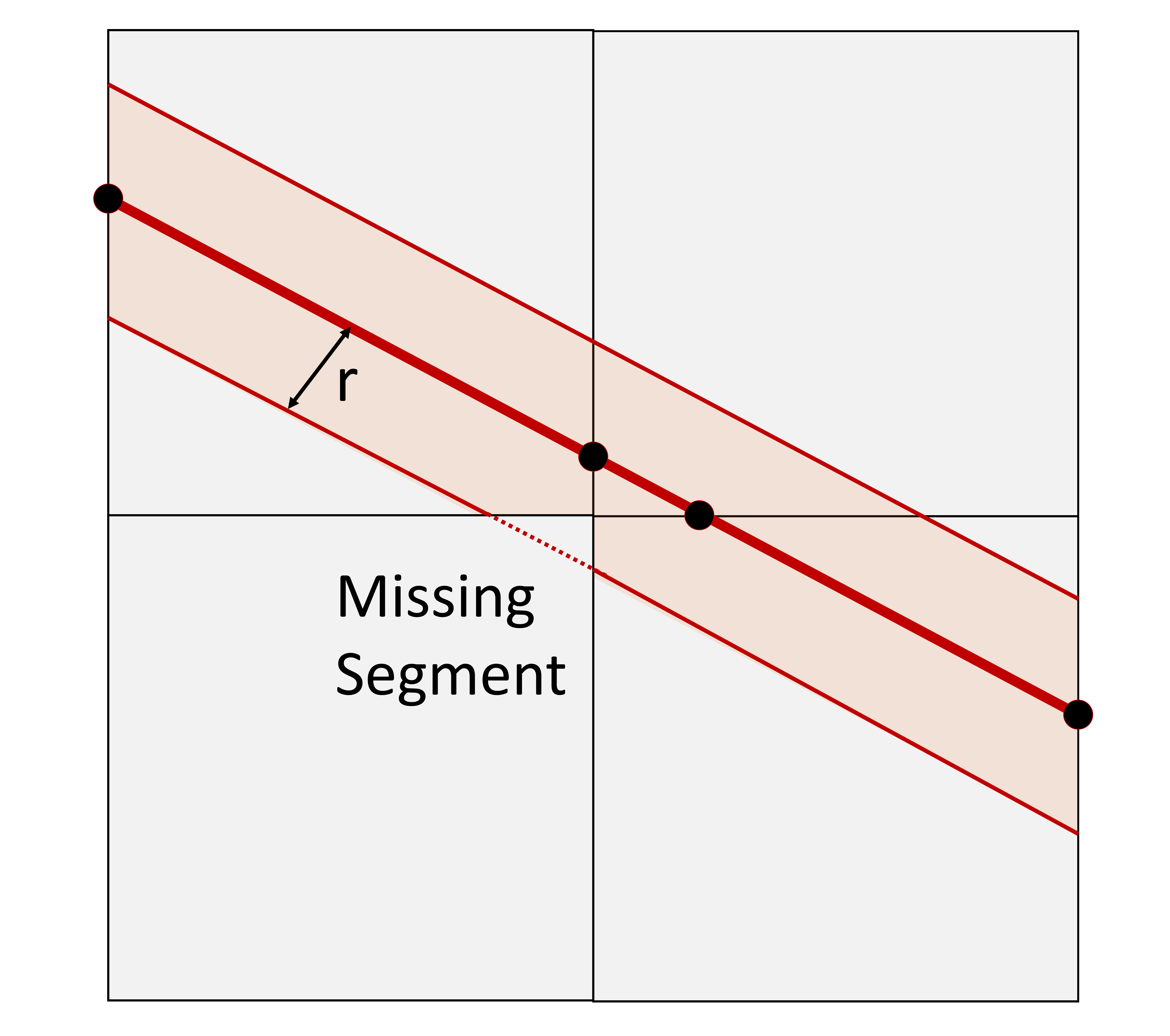}
\includegraphics[width=.23\textwidth]{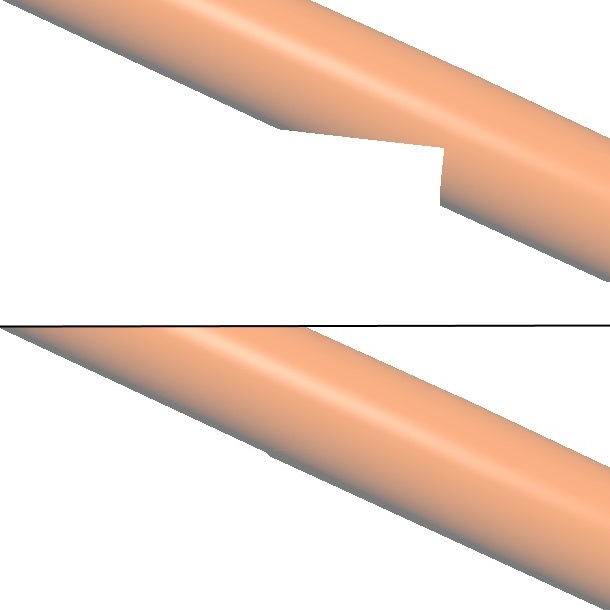}
\caption{When a tube extends into a voxel, and a ray intersects that voxel but none of the voxels in which the tube's center line is encoded (left), no intersection is found and a piece of the tube is missing (right top). By including neighbouring voxels in the intersection test, missing intersection points are found (right bottom).} 
\label{fig:missing_segments}
\end{figure}

During ray-casting, a problematic case can occur if a tube stands out of the voxel in which its center line is defined. Since the tube expands into a voxel which may not know the center line, the piece in this voxel cannot be rendered if the ray doesn't intersect any of the voxels in which the center line is encoded. This situation is illustrated in Fig.~\ref{fig:missing_segments}. 
A straightforward solution to this problem is for a given ray and voxel intersection to also test for intersections with lines from adjacent voxels, a solution that unfortunately decreases rendering performance
due to the increased number of intersection tests. 
However, our experiments have shown that the artifacts that occur when neglecting the missing segments are only rarely visible. In particular when rendering tubes with transparency the artifacts are hardly perceivable. 
As a compromise, we hence during interactive navigation restrict our method to testing against the lines in the voxel hit by the ray. 
As soon as the camera stands still, adjacent voxels are also taken into account.

\begin{figure}[h]
\centering
\includegraphics[width=.23\textwidth]{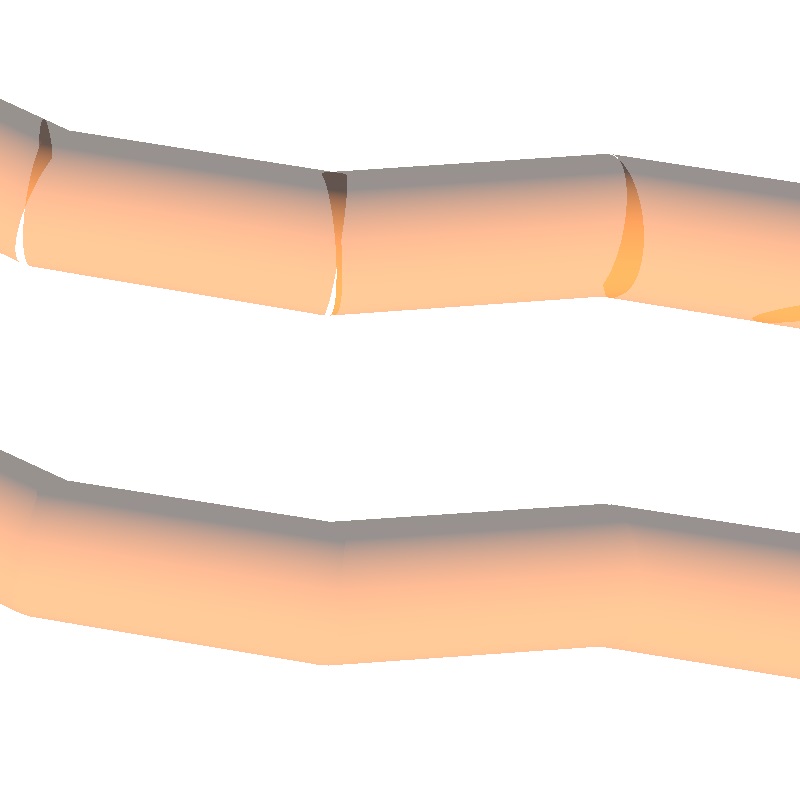}
\includegraphics[width=.23\textwidth]{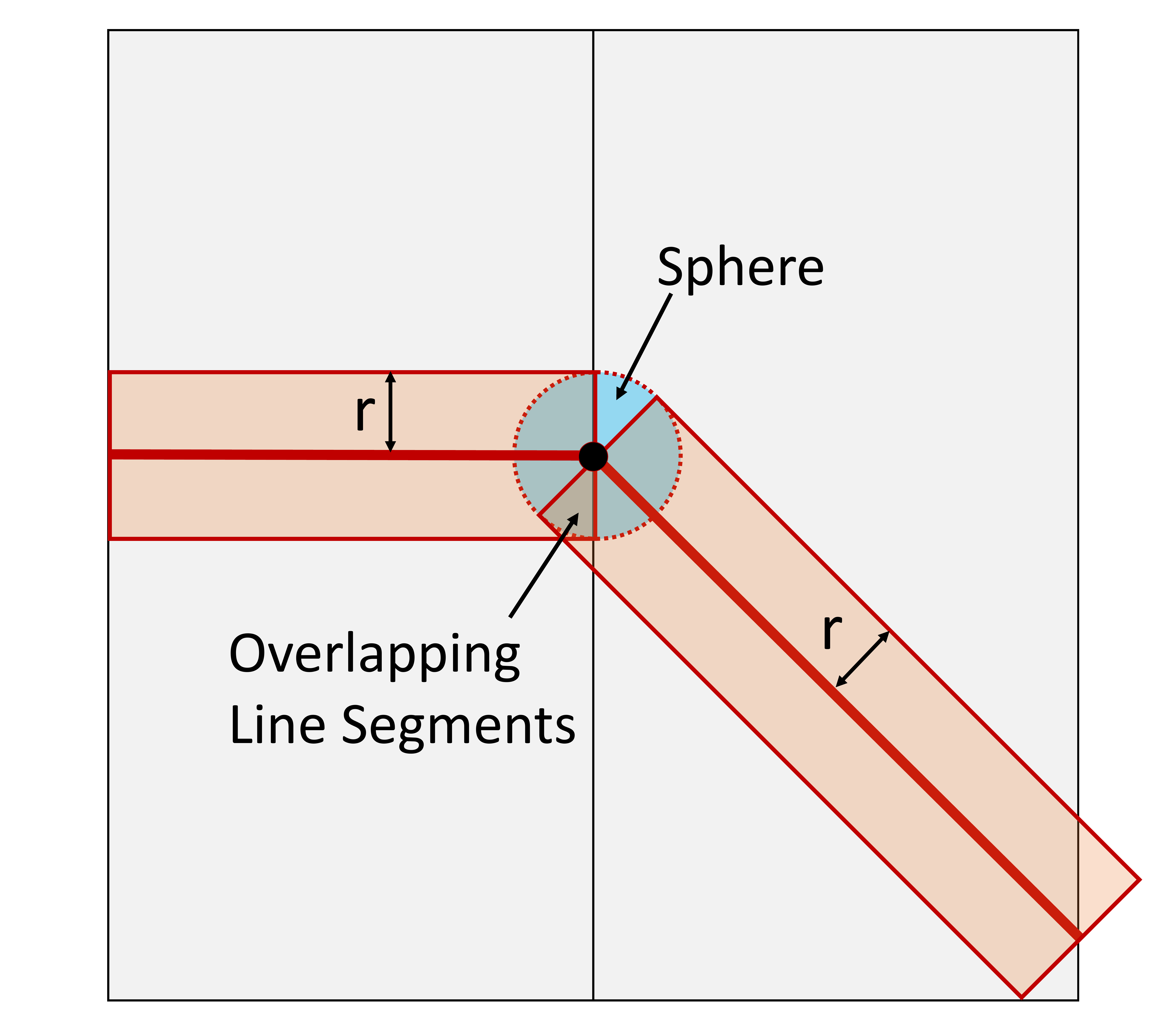}
\caption{Left top: Overlapping tubes are blended incorrectly and parts are missing. Rendering a sphere at line joints (right) continuously closes the gaps (left bottom).}
\label{fig:missing_sphere}
\end{figure}

Another problematic situation occurs at the joint between adjacent lines, i.e., a gap is produced if the lines do not have the same direction (see Fig.~\ref{fig:missing_sphere}). The gap is filled by rendering spheres at the end points of the line segments with a radius identical to the tube's radius. To avoid blending the same line twice, we keep track of the IDs of intersected lines using bit operations and consider an intersection point only once.

\section{Rendering effects}
\label{sec:rendering}
\begin{figure}[h]
\includegraphics[width=0.24\textwidth]{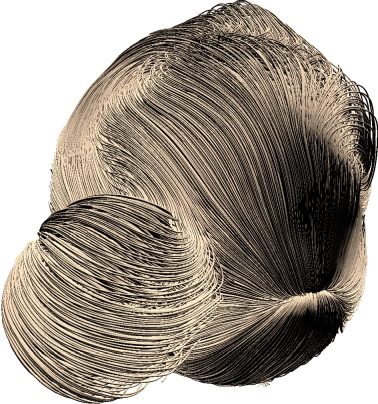}\hfill
\includegraphics[width=0.24\textwidth]{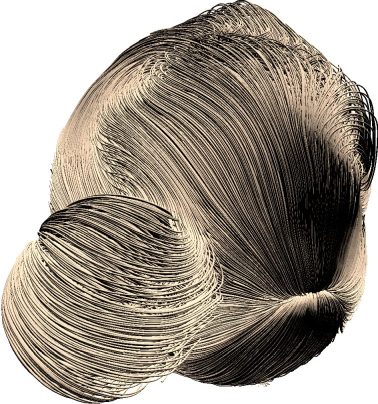}
\caption{Rendering of opaque lines. Left: GPU ray-casting of discretized lines (8ms, $256^3$ voxel grid, curve-face intersections quantized to $16^2$ bins). Right: GPU rasterization of initial lines (12 ms). }
\label{fig:rast-vs-raycast}
\end{figure}

At every entering ray-tube intersection point, we calculate the tube normal and evaluate a local illumination model. If the tubes are rendered opaque, this only has to be done for the first intersection point. The resulting color value is combined with the tube color, and this color is used as pixel color. Fig.~\ref{fig:rast-vs-raycast} demonstrates the rendering of opaque tubes via GPU ray-casting, and compares the result to that of GPU rasterization of the initial lines.

Interestingly, in terms of rendering quality rasterization and ray-casting do not seem to show any significant differences at the selected discretization resolution, yet even for opaque lines ray-casting renders already faster than rasterization. 
The main reason is that ray-casting can effectively employ early-ray termination once the first intersection with a tube is determined. 
Since rasterization renders the lines in the order they are stored, which is  not the visibility order in general, it needs to generate a considerably larger number of fragments. 

\subsection{Transparency rendering}
If the tubes are rendered semi-transparent, at every entering ray-tube intersection an opacity value $\alpha$ is either read from the voxel header or assigned via a transfer function. The opacity value is then used to modulate the tube color, and this color is blended with the pixel color using front-to-back $\alpha$-compositing, i.e., in the order in which the tube intersections are determined along the ray. The described way of handling opacity is exactly how the final pixel color is computed when semi-transparent lines are rendered via GPU rasterization. In contrast to ray-casting, however, in GPU rasterization all generated fragments first have to be stored and finally sorted per-pixel with respect to increasing depth. Only then can the fragments' colors be blended in correct order.

\begin{figure}[h]
\includegraphics[width=0.24\textwidth]{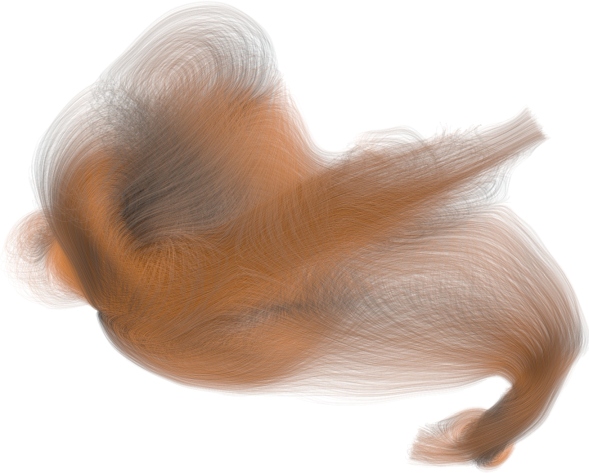}\hfill
\includegraphics[width=0.24\textwidth]{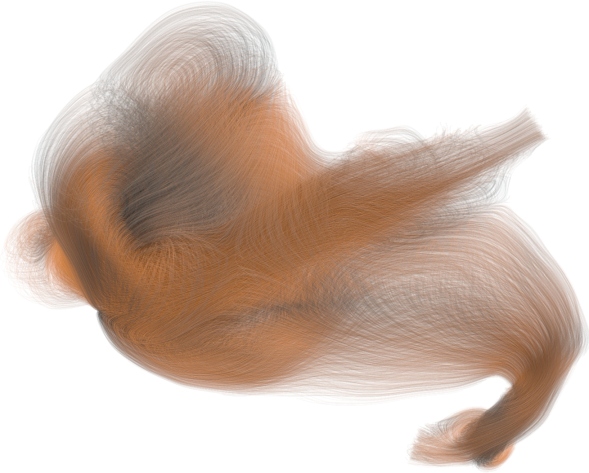}
\caption{Left: Rasterization-based transparency rendering using fragment linked lists on the GPU (160ms, 600 MB fragment list). Right: GPU ray-casting (24ms, $256^3$ voxel grid, curve-face intersections quantized to $16^2$ bins, 60 MB voxel representation).}
\label{fig:LinkedListVsRaycast}
\end{figure}

Fig.~\ref{fig:LinkedListVsRaycast} shows colored and semi-transparent lines, once rendered via GPU rasterization and once via GPU ray-casting. In this situation, the advantage of ray-casting comes out most significantly: Since the ray-tube intersection points are computed in correct visibility order, there is no need to store and finally sort these points for blending. Due to this, the performance gain of ray-casting compared to rasterization now becomes significant; about a factor of 7 for the used dataset.

Furthermore, additional acceleration and quality improvement strategies can be integrated into ray-casting in a straight forward way. Firstly, $\alpha$-termination, i.e., the termination of a ray once the accumulated opacity exceeds a user-defined threshold, can be used to reduce the number of ray-tube intersection points. If a LoD representation is available, even opacity-acceleration\cite{Danskin:1992:FAV:147130.147155} can be employed, i.e., increasing the step size along the ray and simultaneously sampling the opacity from ever coarser resolution levels with increasing optical depth. Secondly, instead of considering a constant opacity per tube, the opacity can be made dependent on the distance the ray travels within the tube. Since together with the ray entry point also the ray exit point is computed, this distance is immediately available. Even the handling of penetrating tubes does not impose any conceptual problem, and only requires to sort the ray-tube intersections locally per voxel.

\subsection{Shadow simulation}

The possibility to efficiently trace arbitrary rays through the voxelized 3D curves can be employed to efficiently simulate global illumination effects such as shadows. Shadows provide additional depth and shape cues, and can significantly enhance the visual perception of the curves geometry and their spatial relationships.

\begin{figure}[h]

\begin{tikzpicture}
\node (imgA)[anchor=south west,inner sep=0] at (0,0)
	{
	\includegraphics[width=0.24\textwidth]{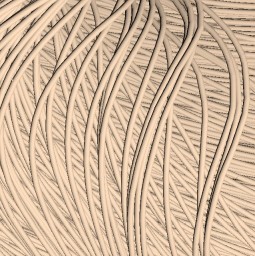}
	\includegraphics[width=0.24\textwidth]{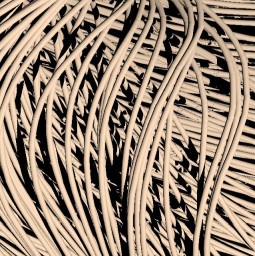}
	};
\node[fill=white, fill opacity=.95, text opacity=1,anchor=south west,inner sep=3pt] at(0.1,0){a)};
\node[fill=white, fill opacity=.95, text opacity=1,anchor=south west,inner sep=3pt] at(4.5,0){b)};
\end{tikzpicture}

\begin{tikzpicture}
\node (imgA)[anchor=south west,inner sep=0] at (0,0)
	{
	\includegraphics[width=0.24\textwidth]{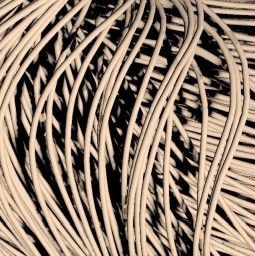}
	\includegraphics[width=0.24\textwidth]{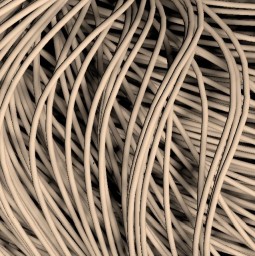}
	};
\node[fill=white, fill opacity=.95, text opacity=1,anchor=south west,inner sep=3pt] at(0.1,0){c)};
\node[fill=white, fill opacity=.95, text opacity=1,anchor=south west,inner sep=3pt] at(4.5,0){d)};
\end{tikzpicture}

\caption{(a) Local illumination. (b) Per-fragment hard shadows. (c) Shadows from representative lines at first LoD. (d) Soft shadows from line density values via cone-tracing.}
\label{fig:softshadows}
\label{fig:hardshadows}
\end{figure}

The simulation of hard shadows of a point light source can be realized by sending out shadow rays and testing if the ray hits another tube before it hits a light source. However, as demonstrated in Fig.~\ref{fig:hardshadows}b, due to the high frequency shadow patterns that are caused by a dense set of curves, hard shadows rather disturb the visual perception than help to improve it.

We propose the following two approaches to incorporate shadows into the rendering of large line sets without introducing high-frequency shadow patterns. The first approach is to test the shadow rays against the representative lines at a coarser LoD, thus reducing the number of lines that throw a shadow and making the shadows wider and more contiguous (see Fig.~\ref{fig:softshadows}c). The second approach replaces hard shadows by soft shadows, by sampling the line density values along the shadow rays to measure the amount of blocking. (see Fig.~\ref{fig:softshadows}d). Both approaches require to traverse only one single shadow ray towards the light source and can be performed efficiently on the proposed LoD representation.

In the first approach, the shadow rays traverse the voxels of a selected LoD and test for intersections with the representative line in every voxel. Even though we do not avoid hard shadows in this way, by using one single yet thicker line to represent many thinner lines, the frequency of variations from shadow to non-shadow can be reduced significantly. The second approach mimics the effect of an area light source, requiring, in general, to use many rays to estimate how much of the light leaving the area light is blocked. Instead, we sample the line density values with one ray in a way similar to cone-tracing~\cite{Crassin:2011:III:1944745.1944787}, i.e., by sampling from ever coarser resolution levels with increasing distance from the illuminated point. In particular, we simulate the amount of light that falls onto the point along a cone with an opening angle that subtends one voxel at the finest level.

When comparing the results of both approaches to the rendering of hard shadows in Fig.~\ref{fig:hardshadows}a, one can see that the shadow frequency is considerably reduced and the visual perception is improved. The major shadowing effects, on the other hand, are still present in the final renderings, and the spatial relationships between the lines are effectively revealed. When using the first approach, the render time increases about $20\%$ compared to the rendering without shadows; the second approach yields a decrease of about $10\%$. These only marginal decreases are due to the use of the LoD representation, which requires testing against only one single line per voxel when using the first approach, and interpolating trilinearly in the line density fields at different resolution levels when using the second approach. Since a texture lookup operation takes less time than an explicit ray-tube intersection test, the second approach performs even faster than the first one at almost similar quality.
\subsection{Ambient occlusion}

\begin{figure}[h]

\begin{overpic}[width=0.24\textwidth]{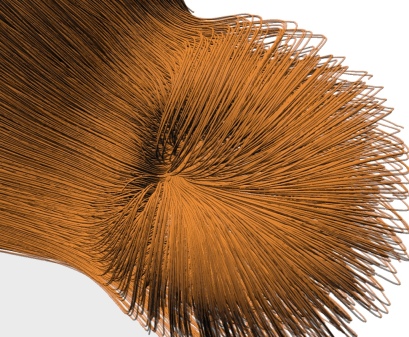}
	\put(4,2){ {\textcolor{black}{a)}}}
\end{overpic}\hfill
\begin{overpic}[width=0.24\textwidth]{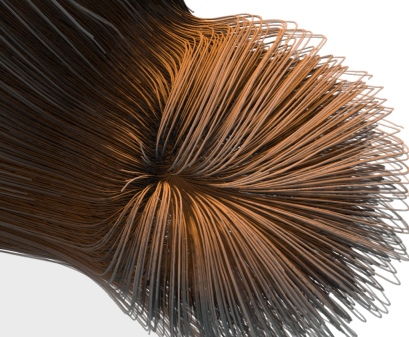}
	\put(4,2){ {\textcolor{black}{b)}}}
\end{overpic}
\caption{Local lighting without (a) and with (b) ambient occlusions.}
\label{fig:ao_comparison}
\end{figure}

Both rendering approaches for low-frequency shadows can also be used to simulate ambient occlusions (AO), i.e., soft shadows that occur in the cavities of a 3D object when indirect lighting is cast out onto the scene.
As demonstrated in Fig.~\ref{fig:ao_comparison}, the soft shadows from ambient occlusions help in particular to enhance the spatial separation between individual lines or bundles of lines.

\begin{figure}[t]

\begin{overpic}[width=0.24\textwidth]{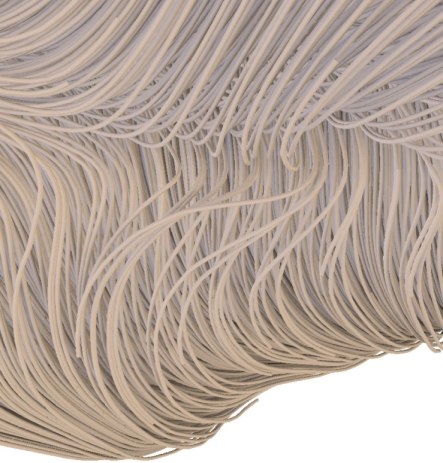}
\put(80,4){ {\textcolor{black}{a)}}}
\end{overpic}\hfill
\begin{overpic}[width=0.24\textwidth]{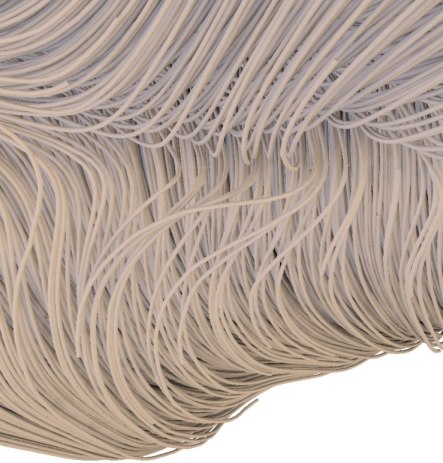}
\put(80,4){ {\textcolor{black}{b)}}}
\end{overpic}\hfill
\begin{overpic}[width=0.24\textwidth]{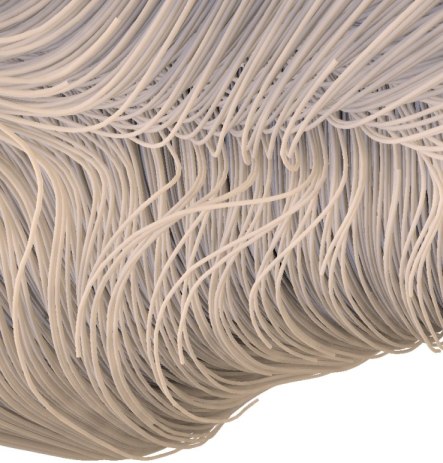}
\put(80,4){ {\textcolor{black}{c)}}}
\end{overpic}\hfill
\begin{overpic}[width=0.24\textwidth]{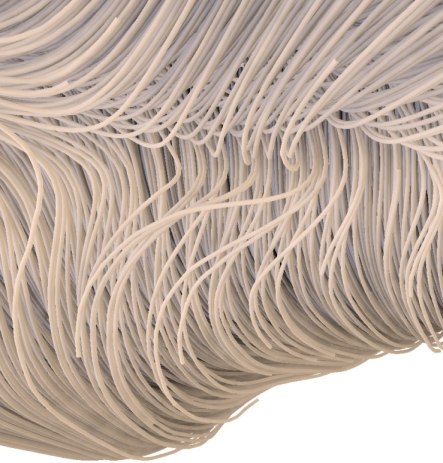}
\put(80,4){ {\textcolor{black}{d)}}}
\end{overpic}\hfill
\caption{(a) Hemisphere ambient occlusion on line geometry. (b) Rays sample the occlusion from the line density values at the first LoD. (c) Spherical occlusion is pre-computed per voxel and sampled trilinearly using 2500 sample rays, respectively (d) 25 sample rays. }
\label{fig:lod_ao}
\end{figure}

AO is simulated by casting out rays to sample the surrounding geometry, and computing how much light from an environment map is blocked by this geometry. For simulating shadows we restrict the direction along which we send out rays to the direction of a directional light or towards a point light source, yet ambient occlusion requires to send out rays into the entire upper hemisphere with respect to the normal direction at a surface point. AO can be integrated in a straight forward way into the ray-based rendering pipeline, by spawning at every visible point a number of secondary rays into the hemisphere and calculating whether the environment light is blocked or not. In Fig.~\ref{fig:lod_ao}a, 2500 rays per visible point were used to uniformly sample the hemisphere, and these rays were intersected against the tube segments in every voxel to determine whether the light is blocked or not. The higher variance in the ambient occlusion values when less rays are used can clearly be seen.

\newcommand{\VariableWidthDatasets}{.2\textwidth}

\begin{figure*}[ht]
\includegraphics[width=\VariableWidthDatasets]{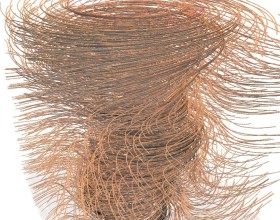}\hfill
\includegraphics[width=\VariableWidthDatasets]{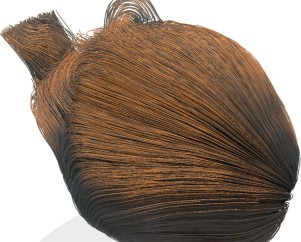}\hfill
\includegraphics[width=\VariableWidthDatasets]{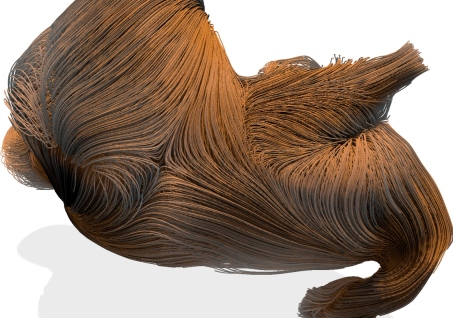}\hfill
\includegraphics[width=\VariableWidthDatasets]{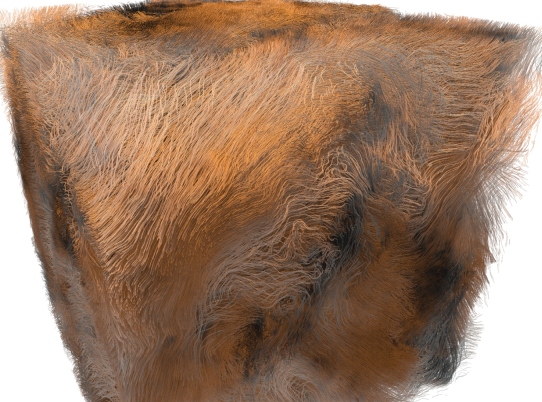}\hfill
\includegraphics[width=\VariableWidthDatasets]{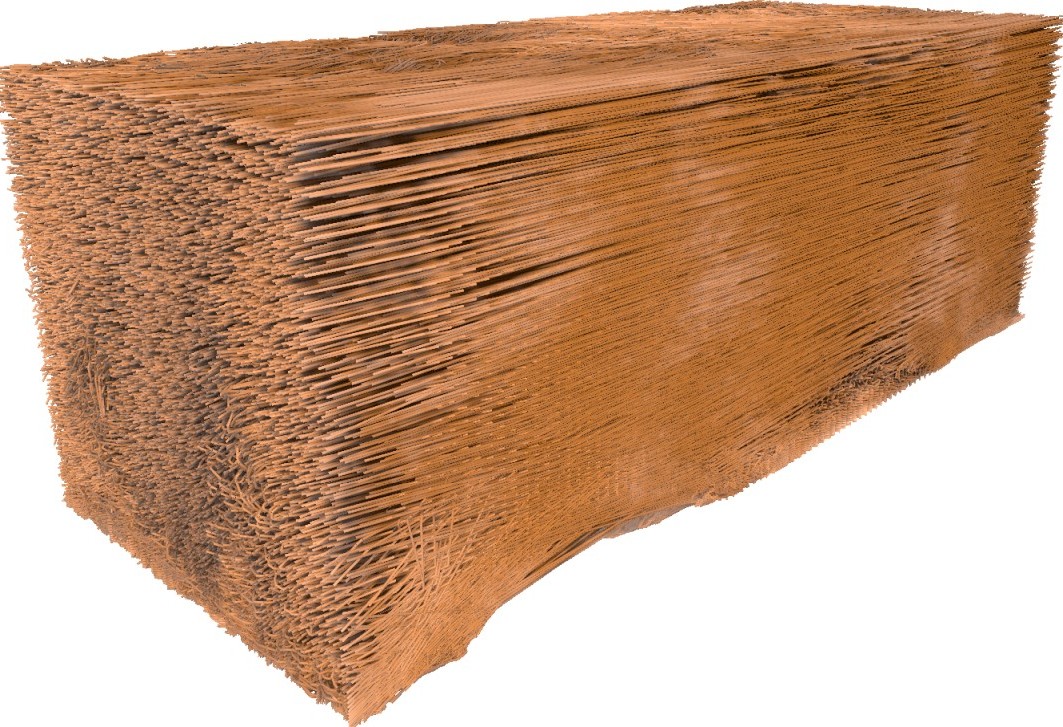}\hfill
\caption{The datasets we have used in our experiments: Tornado, Aneurysm I, Aneurysm II, Turbulence, Weather Forecast.}
\label{fig:datasets}
\end{figure*}
To enable interactive updates of AO values when the scene or light situation changes, the number of rays as well as the number of objects against which the shadow rays are tested need to be reduced.
Screen-space approaches~\cite{Bavoil:2008:IHA:1401032.1401061} compute a rough AO estimation by using a few rays in 2D screen-space (between 8 and 20 in realtime applications), and by testing these rays against rendered surface points in a short radius of influence.

To overcome visual shortcomings of screen-space calculation, we compute AO values in 3D space, and present two acceleration strategies to efficiently approximate the amount of occlusion per tube point. Similar to screen-space ambient occlusion, we restrict the sampling of occluding structures to a radius of influence (in our case 15 voxels at the finest voxel resolution).

The first approach is to approximate the amount of occlusion along a ray by sampling the line density values at the finest level voxel grid via trilinear interpolation (Fig.~\ref{fig:lod_ao}b), instead of testing against the tube geometries at this level (Fig.~\ref{fig:lod_ao}a). The values along a ray are accumulated until either full blocking is reached (line density $\geq$ 1) or the ray reaches beyond the radius of influence of leaves the domain. The occlusion values are finally integrated over all rays and normalized by dividing through the number of rays.

The renderings in Figs.~\ref{fig:lod_ao}a and b show that the AO values vary strongly around the tube axis, which is due the hemisphere sampling of occlusions wrt to the varying normal direction. Due to the many tubes that are rendered, this adds high frequent intensity variations which rather disturb the visual impression than help to enhance the spatial relationships between the tubes. To avoid this effect, we propose a second approach which computes point-wise AO values independent of the surface orientation by considering occlusions in the entire sphere around each visible point (see Fig.~\ref{fig:lod_ao}c and d). The AO values are first approximated per voxel in a pre-process, and at runtime these values are trilinearly interpolated at the locations of the visible tube points. In this way the AO values reflect the local spherical surrounding of a line rather than the surrounding in the normal direction at the tube points. This emphasizes in a far better way the embedding of a line in the surrounding set of lines.

%% file: sections/results.tex
\section{Results and Discussion}
\label{sec:results}

In this section, we analyze the quality, memory consumption, and performance of our approach. All times were measured on a standard desktop PC, equipped with an Intel Xeon E5-1650 v3 CPU with 6$\times$3.50~GHz, 32~GB RAM, and an NVIDIA GeForce GTX 970 graphics card with 4~GB VRAM. For all renderings, the view port was set to 1920$\times$1080.

We used the following datasets to analyze the quality, memory requirements, and performance of the proposed voxel-based rendering approach for line sets (see Fig.~\ref{fig:datasets}):
\begin{itemize}
\item \textbf{Tornado}: 1000 randomly seeded streamlines in a flow forming a tornado.  The lines have a higher curvature as well as a higher density at the center of the tornado.
\item \textbf{Aneurysm I\,/\,II}: 4700\,/\,9200 randomly seeded streamlines in the interior of two different aneurysms, allowing hemodynamic analysis as done by Byrne et al.~\cite{Byrne13}. The streamlines were advected up to the vascular wall, resulting in empty space up to the cuboid domain boundaries.
\item \textbf{Turbulence}: 50000 domain-filling streamlines advected in a forced turbulence field of resolution $1024^{3}$ as described by Aluie et al.~\cite{Hopkins}.
\item \textbf{Weather Forecast}: 212000 domain-filling path lines computed over 96 hours each on the wind field of a forecast by the European Centre for Medium-Range Weather Forecasts (ECMWF). The path lines were seeded on a horizontally regular grid and on vertically stretched levels\cite{RautenhausEtAl2015GMDb}.
\end{itemize}

\subsection{Quality analysis}

To analyze the effect of the resolution of the voxel grid and the quantization structure on reconstruction quality, we performed a number of experiments with the datasets listed above. 
The subjective visual quality of the rendered images is hard to measure, wherefore we attempt to quantify the quality objectively by measuring the following quantities:
For different resolutions we measured the mean Hausdorff distance between the original curves and their piecewise linear approximations (Fig.~\ref{dia:ErrorPlots}a), the mean angle between the tangents at the original curve points and the curve's piecewise linear approximations (Fig.~\ref{dia:ErrorPlots}b), and the number of line segments falling onto each other due to the quantization of vertex coordinates (Fig.~\ref{dia:ErrorPlots}c).

Fig.~\ref{dia:ErrorPlots}a indicates that already at a voxel grid resolution of $128^3$ and a quantization resolution of 32, suitable reconstruction accuracy is achieved. 
At a voxel grid resolution of $64^3$, geometric features get lost and cannot be faithfully reconstructed even at high quantization resolution. Fig.~\ref{fig:curve_discretization}b shows essentially the same dependencies, yet one can observe a stronger effect of the quantization resolution. The local directional changes of the curve tangents due to displacements of the vertex coordinates is scale independent and depends mainly on the quantization resolution. From Fig.~\ref{fig:curve_discretization}c it can be seen that at a voxel grid resolution of $128^3$ and higher, and starting at a quantization resolution of 32, the number of duplicate lines is below 0.1$\%$ of all encoded lines and doesn't change significantly beyond these resolutions. The maximum Hausdorff distance, on the other hand, is always bounded by the voxel diagonal, and the directional error can be up to $180$ degrees, if a curve makes a loop in a voxel.

Supported by our analysis, and further verified by comparing the visual quality of the original curves and the voxel-based curve representations, we found that a voxel grid resolution of $256^3$ and a quantization level of 32 is for all cases the resolution at which a further increase in voxel resolution or bin size only causes little improvement in error quantities 
This is also evidenced by the close-up views in Fig.~\ref{fig:colorpage-rastray}, where no apparent differences between the original curves and their voxelized counterparts can be perceived. Due to this, we decided to use these resolutions in the performance analysis and the analysis of illumination effects below.

\definecolor{color1}{rgb}{0,0.2,0.9}
\definecolor{color2}{rgb}{1,0.1,0}
\definecolor{color3}{rgb}{1,0.6,0} \definecolor{color4}{rgb}{0.06,0.6,0.1}
\definecolor{color5}{rgb}{   0, 0.15  ,0.6}\definecolor{color6}{rgb}{0.75 , 0.07 ,0}\definecolor{color7}{rgb}{0.75 , 0.45  ,0}\definecolor{color8}{rgb}{0.04, 0.5 ,0.07}
\pgfplotsset{
    cycle list={color1\\color2\\color3\\color4\\color5\\},
}

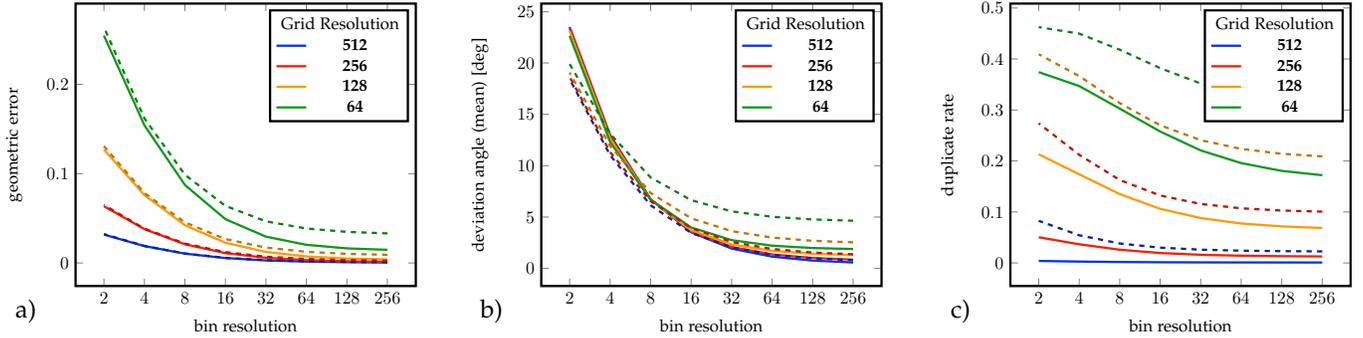
\begin{figure*}
    \centering
 \begin{minipage}[t]{.32\linewidth}
     \centering
	\begin{tikzpicture} [scale = 0.66]
		\begin{axis}[			xmode=log,	
			log basis x=2,
			xticklabel=\pgfmathparse{2^\tick}\pgfmathprintnumber{\pgfmathresult},
			xlabel={bin resolution},						ylabel={geometric error 
									},			line width=1.3pt,
			]

			\addlegendimage{empty legend}
			\addlegendentry{\hspace{-.6cm}Grid Resolution}
			\addlegendentry{\textbf{512}}
			\addlegendentry{\textbf{256}}
			\addlegendentry{\textbf{128}}
			\addlegendentry{\textbf{64}}
			
			\addplot table[x=bins, y=a] {data/aneurysm2/DistanceError.csv};
			\addplot table[x=bins, y=b] {data/aneurysm2/DistanceError.csv};
			\addplot table[x=bins, y=c] {data/aneurysm2/DistanceError.csv};
			\addplot table[x=bins, y=d] {data/aneurysm2/DistanceError.csv};

			\addplot[dashed,color5] table[x=bins, y=a] {data/weather/DistanceError.csv};
			\addplot[dashed,color6]  table[x=bins, y=b] {data/weather/DistanceError.csv};
			\addplot[dashed,color7]  table[x=bins, y=c] {data/weather/DistanceError.csv};
			\addplot[dashed,color8]  table[x=bins, y=d] {data/weather/DistanceError.csv};

			\end{axis}
			\node[] at (-1,-0.5) {a)};
	\end{tikzpicture}
\label{dia:MeanDistanceError}
\end{minipage}\hfill\begin{minipage}[t]{.32\linewidth}
    \centering
	\begin{tikzpicture} [scale = 0.66]
		\begin{axis}[			xmode=log,	
			log basis x=2,
			xticklabel=\pgfmathparse{2^\tick}\pgfmathprintnumber{\pgfmathresult},
			xlabel={bin resolution},			ylabel={deviation angle (mean) [deg]},			line width=1.3pt,
			scaled ticks=false,
			tick label style={/pgf/number format/fixed}
			]

			\addlegendimage{empty legend}
			\addlegendentry{\hspace{-.6cm}Grid Resolution}
			\addlegendentry{\textbf{512}}
			\addlegendentry{\textbf{256}}
			\addlegendentry{\textbf{128}}
			\addlegendentry{\textbf{64}}
			
			\addplot table[x=bins, y=a] {data/aneurysm2/TangentError.csv};
			\addplot table[x=bins, y=b] {data/aneurysm2/TangentError.csv};
			\addplot table[x=bins, y=c] {data/aneurysm2/TangentError.csv};
			\addplot table[x=bins, y=d] {data/aneurysm2/TangentError.csv};

			\addplot[dashed,color5] table[x=bins, y=a] {data/weather/TangentError.csv};
			\addplot[dashed,color6]  table[x=bins, y=b] {data/weather/TangentError.csv};
			\addplot[dashed,color7]  table[x=bins, y=c] {data/weather/TangentError.csv};
			\addplot[dashed,color8]  table[x=bins, y=d] {data/weather/TangentError.csv};

			\end{axis}
			\node[] at (-1,-0.5) {b)};
	\end{tikzpicture}
\label{dia:MeanTangentError}
\end{minipage}\hfill\begin{minipage}[t]{.32\linewidth}
    \centering
	\begin{tikzpicture} [scale = 0.66]
		\begin{axis}[			xmode=log,	
			log basis x=2,
			xticklabel=\pgfmathparse{2^\tick}\pgfmathprintnumber{\pgfmathresult},
			xlabel={bin resolution},			ylabel={duplicate rate},			line width=1.3pt,
			scaled ticks=false,
			]

			\addlegendimage{empty legend}
			\addlegendentry{\hspace{-.6cm}Grid Resolution}
			\addlegendentry{\textbf{512}}
			\addlegendentry{\textbf{256}}
			\addlegendentry{\textbf{128}}
			\addlegendentry{\textbf{64}}
			
			\addplot table[x=bins, y=a] {data/aneurysm2/Duplicates.csv};
			\addplot table[x=bins, y=b] {data/aneurysm2/Duplicates.csv};
			\addplot table[x=bins, y=c] {data/aneurysm2/Duplicates.csv};
			\addplot table[x=bins, y=d] {data/aneurysm2/Duplicates.csv};
			
			\addplot[dashed,color5] table[x=bins, y=a] {data/weather/Duplicates.csv};
			\addplot[dashed,color6]  table[x=bins, y=b] {data/weather/Duplicates.csv};
			\addplot[dashed,color7]  table[x=bins, y=c] {data/weather/Duplicates.csv};
			\addplot[dashed,color8]  table[x=bins, y=d] {data/weather/Duplicates.csv};
			
			\end{axis}
		\node[] at (-1,-0.5) {c)};
	\end{tikzpicture}

\label{dia:Duplicates}
\end{minipage}
\caption{Comparing the Aneurysm II (solid line) with the Weather Forecast (dashed line) dataset using different metrics: (a)~Distance Error (in units of voxel size at a grid of $64^3$), (b)~tangent deviation, (c)~Number of duplicates in proportion to the number of generated segments}
\label{dia:ErrorPlots}
\end{figure*}

\subsection{Memory statistics}

Our approach requires a 5 byte header for every voxel, to indicate how many lines are encoded per voxel (1 byte, restricting to a maximum of $2^8-1$ lines per voxel) and to reference the memory address where the lines are stored (4 bytes). In addition, every line is encoded by specifying at which of the 6 voxel faces the two endpoints are located ($2 \cdot 3$ bits), and addressing the sub-face on each voxel face to which the endpoints are quantized ($2 \cdot 10$ bits for a quantization resolution of $32^2$). Furthermore, we use 1 byte per line to map to a transparency or color value and 5 bits to encode a local line ID. For a selected quantization resolution, the minimum number of bytes that is required to encode this information is used. Overall, our approach requires 4, 5, and 6 bytes per line, respectively, when quantization resolutions of (4,8), (16,32) and (64,128) are used.

Table~\ref{tab:memory} compares the memory consumption for different resolution and quantization levels with the memory that is required to store the original lines on the GPU, using 32 bit float values per vertex component and 1 byte per vertex to map to transparency or color. In the second column we show in brackets the memory that is required in average over multiple views to store the per-pixel fragment lists when transparent lines are rendered. It can be seen that for all but the Tornado dataset the voxel-based representation at our selected resolution level ($256^3/32^3$) has a significantly lower memory footprint. Even when opaque lines are rendered and the extra memory for storing the fragment lists is not required, a rather moderate increase of about a factor of 4 to 5 is observed. One exception is the Weather Forecast dataset, which is comprised of many curves with only few long lines. These lines are split into many smaller lines and stored in the voxel representation, so that the memory requirement is significantly increased compared to the original line representation.

\begin{table*}

  \caption{Memory consumption in MB and preprocessing times in seconds for the used datasets. The original geometry is encoded in three float values per vertex. In the second column, we give in brackets the memory that is required to store the per-pixel fragment lists on the GPU. The remaining columns show the memory that is used by the voxel-based representation---with voxel grid resolution $V^3$ and quantization resolution $Q^2$ given as $V^3/Q^2$---, and in brackets the preprocessing time to generate the representations. Per vertex and per quantized line a 1 byte index is stored.}

  \scriptsize
  \newcommand{\z}{\phantom{0}}
  \begin{center}
    \begin{tabular}{rc|c|ccccccccc}

			& Lines			 		&Original Geometry		& $512^3/256^2$ 	& $256^3/128^2$ 	&$256^3/32^2$	& $128^3/128^2$ 	& $128^3/32^2$ 	& $64^3/8^2$ 	 \\
	Dataset	&(vertices per line)	&(Linked List)		&  (Preprocess)	& 	&  	& 	 	& 	 	&	 	\\
    \hline
	   Tornado   	 & \z1000 (250)		& 3 (100) 			&	811		(1.4)		&	103		(0.2)		&	103		(0.2)	&	14	(0.1)	&	14	(0.1)	&	2	(0.1)	\\
       Aneurysm I	&  4700 (410)		& 25 (600) 			&	731		(1.6)		&	111		(0.4)		&	107		(0.4)	&	25	(0.2)	&	22	(0.2)	&	6	(0.3)	\\
       Aneurysm II	&  9200 (367)		& 44 (750)			&	702		(1.9)		&	116		(0.6)		&	109		(0.6)	&	29	(0.5)	&	26	(0.5)	&	8	(0.4)	\\
       Turbulence	&  50000 (220)		& 143 (2500) 		&	1087	(3.5)		&	221		(1.4)		&	201		(1.4)	&	72	(1.3)	&	62	(1.3)	&	21	(1.0)	\\
       Weather Forecast&  212000 (13)	& 36 ($>$4000)		&	573		(5.6)		&	209		(4.6)		&	177		(4.6)	&	98	(5.0)	&	82	(5.0)	&	31	(4.7)	\\
    \end{tabular}
  \end{center}
  \label{tab:memory}
\end{table*}

It is in particular interesting that due to memory limitations on the GPU the Weather Forecast dataset cannot be rendered via GPU rasterization if transparency is used (see Fig. \ref{fig:out-of-memory}). In the figure, red pixels indicate that more than 512 fragments fall onto these pixels, while pixels not colored indicate that not all fragments can be stored in the per-pixel fragment lists due to limited GPU memory. Our approach, in contrast, requires only 144 MB to store the view-independent voxel-based representation on the GPU.

\newcommand{\VariableWidthAA}{.24\textwidth}

\begin{figure}[h]
\includegraphics[width=\VariableWidthAA]{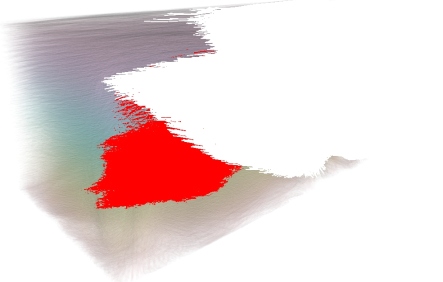}\hfill
\includegraphics[width=\VariableWidthAA]{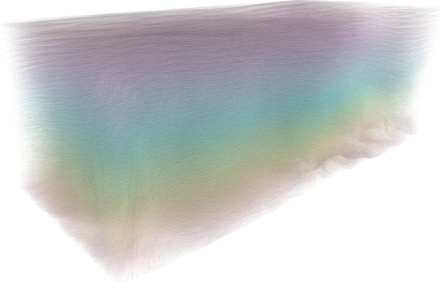}
\caption{Left: GPU rendering using fragment linked lists fails due to memory limitations. a red pixel indicates that more than 512 fragments fall onto that pixel, a pixel remains white when not all fragments could be stored due to memory limitations. Right: The voxel-based representation can be rendered at 23 ms and 144 ms, respectively, when early-ray termination due to accumulated opacity exceeding 0.95 was turned on and off.}
\label{fig:out-of-memory}
\end{figure}

\subsection{Performance analysis}
\label{sec:performance}

We compare the performance of rasterization-based line rendering with and without fragment lists and voxel-based ray-casting for all datasets (see Tab.~\ref{tab:timings}). Images showing the use of transparency to reveal interior structures that are occluded when opaque lines are rendered are shown in the first column of Fig.~\ref{fig:colorpage-light}. Even when opaque lines are rendered, so that fragment lists and sorting is not required in GPU rasterization, the larger datasets can be rendered at higher rates using voxel-based ray-casting. The main reason is that a ray can be terminated immediately when the first ray-tube intersection is computed, while GPU rasterization always needs to generate all fragments even if the early depth-test can discard many of them before entering the fragment stage. On the other hand, for the Tornado dataset, where the line density is rather low so that many rays need to be traversed through the entire voxel grid, voxel-based ray-casting performs slower than rasterization-based rendering. This still holds when transparency is used. The fact that in rasterization-based rendering using transparency all fragments need to be sorted explains the over-linear increase of the render times, while the render times remain almost constant when ray-casting is used on the voxel-based representation.

The worst case scenario for the ray-caster is a very dense line set with very low line opacity. This prevents rays from terminating early, so that all lines along the rays have to be accessed and tested for intersections. Even though this scenario is quite unusual, we analyzed the rendering performance in this situation for the Weather Forecast and Turbulence dataset. In this case, the Forecast dataset can still be ray-cast in roughly 144 ms, while the dataset cannot be rendered via rasterization because the fragment lists exceed the available GPU memory. When using the Turbulence dataset with very low line opacity, GPU rasterization and voxel-based ray-casting render at 380 ms and 124 ms, respectively, demonstrating the efficiency or our approach even in this extreme situation.

\begin{table}[h]
 \caption{Performance statistics for different datasets. Rendering performance was evaluated using a constant per-line opacity value of 25 percent. }
  \scriptsize
  \newcommand{\z}{\phantom{0}}
  \begin{center}
    \begin{tabular}{r|cc|cc}
    	                              & \multicolumn{2}{c}{Timings Semi Opaque} & \multicolumn{2}{c}{Timings Opaque} \\
	Dataset &Linked List&Raycasting& Rasterization&Raycasting\\
    \hline
	   Tornado          &    4 ms	&15 ms			&	2 ms	&	12 ms	\\
       Aneurysm I       &     46 ms	&25 ms		&	8 ms	&	9 ms	\\
       Aneurysm II      &  160 ms   &27 ms		&	14 ms	&	12 ms	\\
       Turbulence       &  380 ms   &24 ms		&	65 ms	&	6 ms	\\
       Weather Forecast &  overflow   & 23 ms	&	29 ms	&	4 ms	\\
    \end{tabular}
  \end{center}
  \label{tab:timings}
\end{table}
\vspace{-0.5cm}
\subsection{Illumination effects} 
The secondary rays used to simulate global illumination effects like shadows and ambient occlusions are solely traversed on the first LoD level of the voxel grid. Therefore, these effects cause only a moderate increase in the render time, yet they can significantly improve the spatial perception of the rendered line structures. This is demonstrated in Fig.~\ref{fig:colorpage-light}, where opaque lines are rendered with soft shadows and ambient occlusions (right).

For instance, for 100 rays sampling in a sphere with a radius of roughly 5 times the voxel size (with a step size of 1 voxel size), ambient occlusions for all $256^3$ voxels can be computed in roughly 9 ms.
Soft shadows can be computed in about 2 ms. The computed illumination values can be stored inside our voxel representation and only have to be updated if the scene changes.

Fig.~\ref{fig:colorpage-rastray} shows a comparison between reference images produced by a rasterizer using fragment linked lists (left) and images produced using our ray-casting approach using a memory efficient representation (right). In all images an attribute stored for each line segment was mapped by a transfer function to a color value. Opacity values were manually predefined.
The memory efficient representation introduces some artifacts like jittered line segments and visible gradations at the voxel boundaries as attributes are not interpolated in between. Comparing the reference image and the high resolution represenation, no significant differences are observable.

In Fig.~\ref{fig:colorpage-light} we compare local illumination (left) and soft shadows in combination with ambient occlusions (right). Semi opaque renderings of high numbers of curves tend to lose sharpness due to overlaying. Ambient occlusion effects are helping to differentiate the individual curves as less light is reaching covered curves in the back. Shadow effects are enhancing the structure and illustrate the relative position of curves to each other. In combination with ambient occlusion, details in shadowed regions are preserved.

%% file: sections/conclusion.tex
\section{Conclusion and future work}
\label{sec:conclusion}

We have proposed a new approach for visualizing large 3D line sets on the GPU, by performing parallel ray-casting of lines on a novel voxel representation for lines. For large line sets we have shown significant improvements over rasterization-based approaches, in terms of rendering performance and quality; by enabling the efficient realization of additional visualization options like global illumination effects. By means of our approach, even very large and complicated line sets can be analysed interactively. 

A limitation of the ray-casting approach for lines is the lack of hardware accelerated anti-aliasing. When using rasterization, anti-aliasing doesn’t affect the performance a lot, yet anti-aliasing in ray-casting is expensive because several rays have to be evaluated per pixel. Another limitation is with respect to the addition of new lines. Since on the GPU the lines are tightly stored in one linear array, the whole memory needs to be restructured when new lines are to be added. On the other hand, generating the voxelization is sufficiently fast, so that a complete re-voxelization of all and the new lines can be done.

In future work we will examine the extension of our approach for the efficient handling of update operations and time-varying line sets. Even though the GPU implementation is almost fast enough to generate the voxel model from scratch if changes occur, local update operations and encodings of the dynamics of lines into the voxel structure will be favorable.

%% file: sections/colorpage.tex
\newcommand{\VariableWidthIII}{.33\textwidth}
\newcommand{\VariableWidthIIII}{.24\textwidth}
\newcommand{\VariableWidthII}{.49\textwidth}

\begin{figure*}

\includegraphics[width=\VariableWidthIIII]{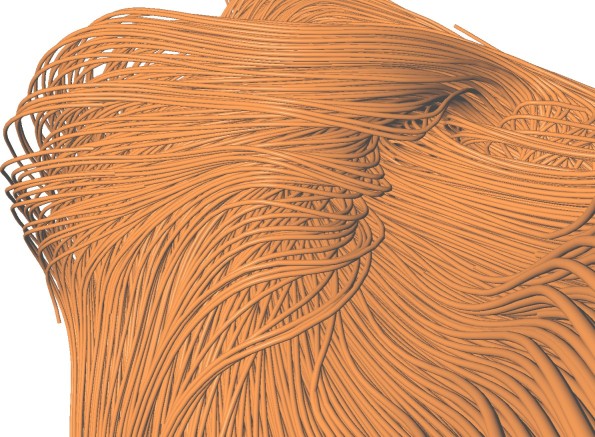}\hfill
\includegraphics[width=\VariableWidthIIII]{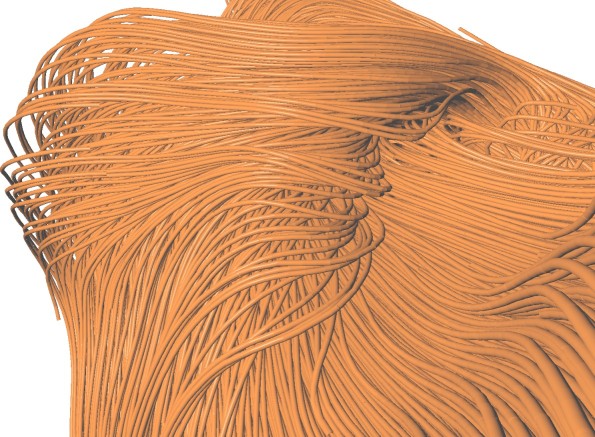}\hfill
\includegraphics[width=\VariableWidthIIII]{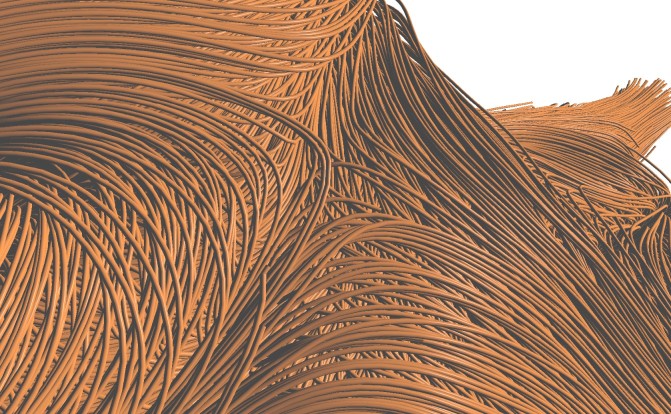}\hfill
\includegraphics[width=\VariableWidthIIII]{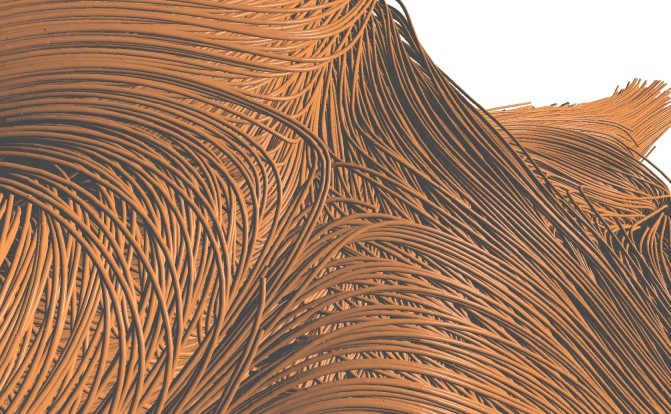}

\includegraphics[width=\VariableWidthIIII]{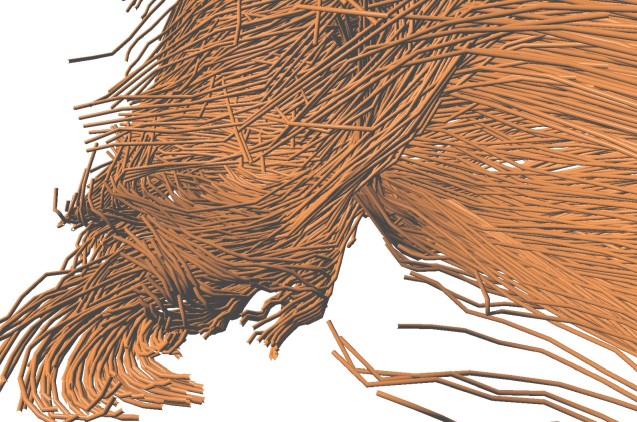}\hfill
\includegraphics[width=\VariableWidthIIII]{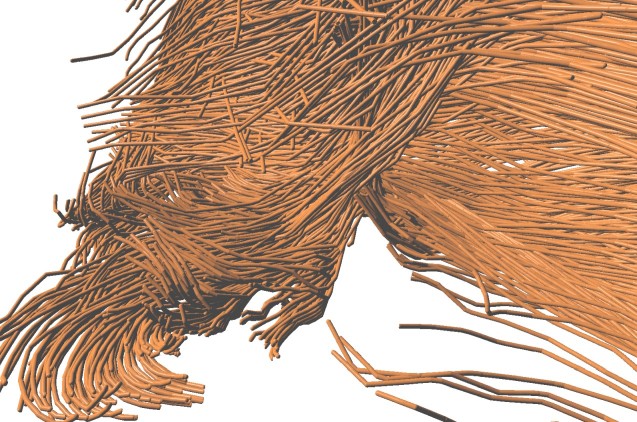}\hfill
\includegraphics[width=\VariableWidthIIII]{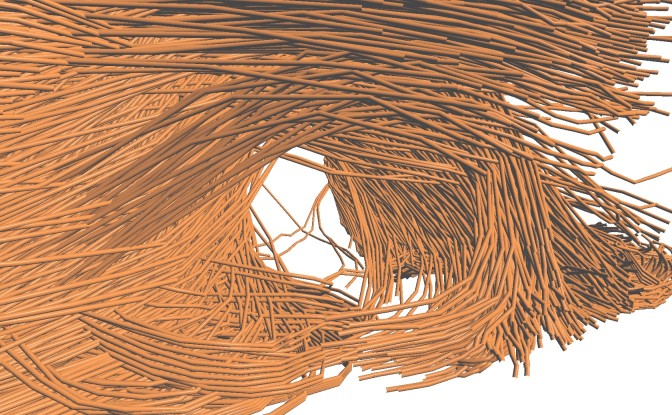}\hfill
\includegraphics[width=\VariableWidthIIII]{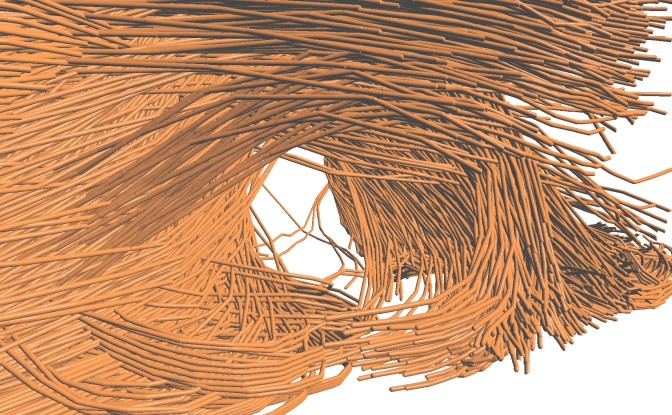}

\caption{Comparison between rasterization-based rendering (left) and voxel-based line raycasting using $256^3$ voxels partitioned into $32^2$ bins per face (right), for the Aneurysm II and the Weather Forecast dataset.
}
\label{fig:colorpage-rastray}
\end{figure*}

\begin{figure*}

\includegraphics[width=\VariableWidthIIII]{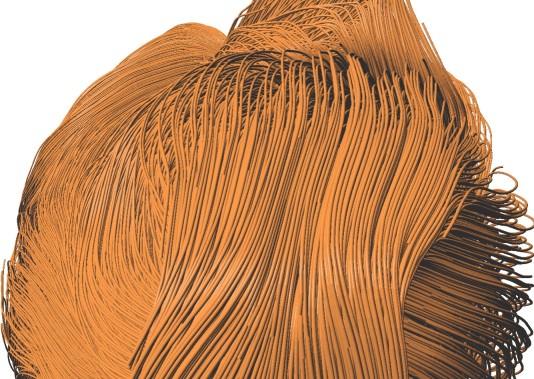}\hfill
\includegraphics[width=\VariableWidthIIII]{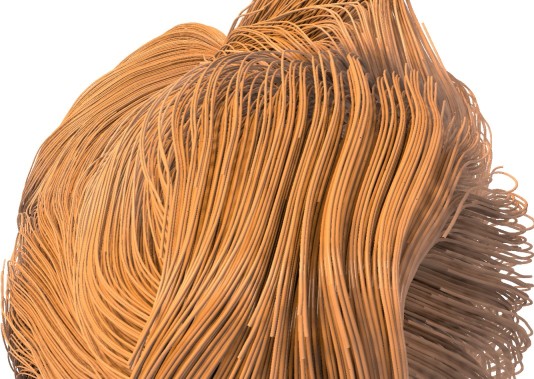}\hfill
\includegraphics[width=\VariableWidthIIII]{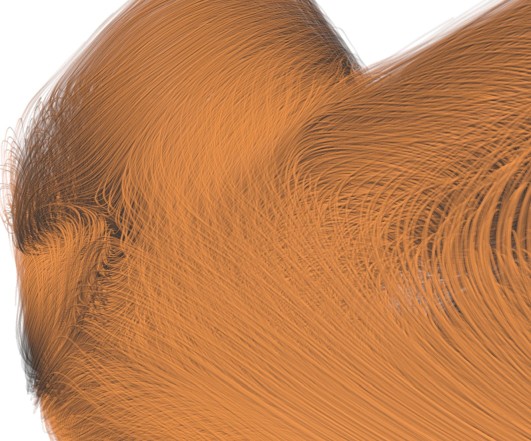}\hfill
\includegraphics[width=\VariableWidthIIII]{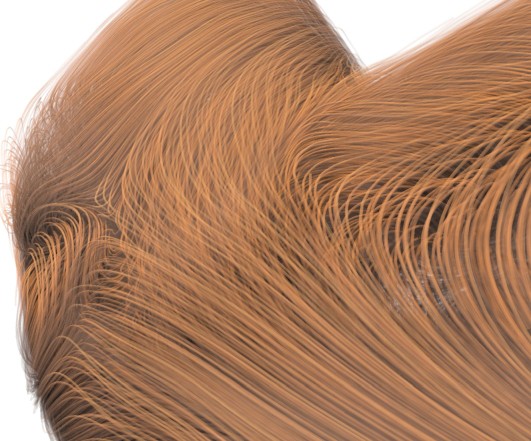}

\includegraphics[width=\VariableWidthIIII]{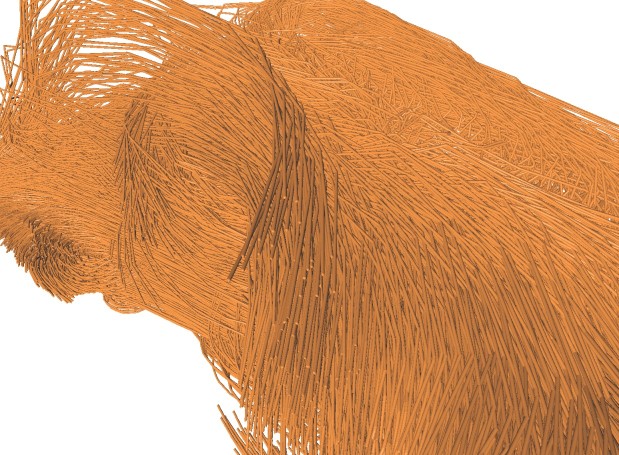}\hfill
\includegraphics[width=\VariableWidthIIII]{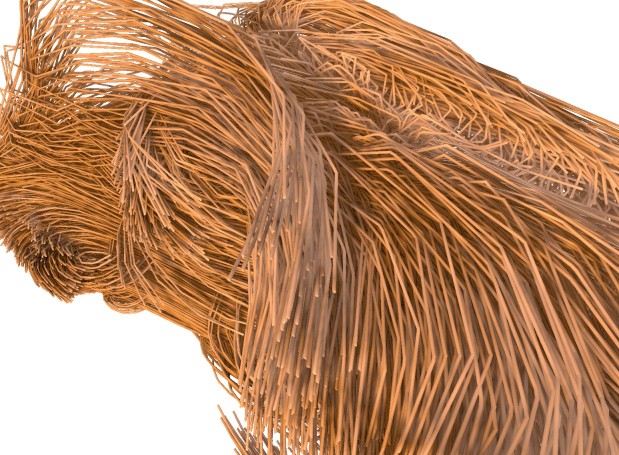}\hfill
\includegraphics[width=\VariableWidthIIII]{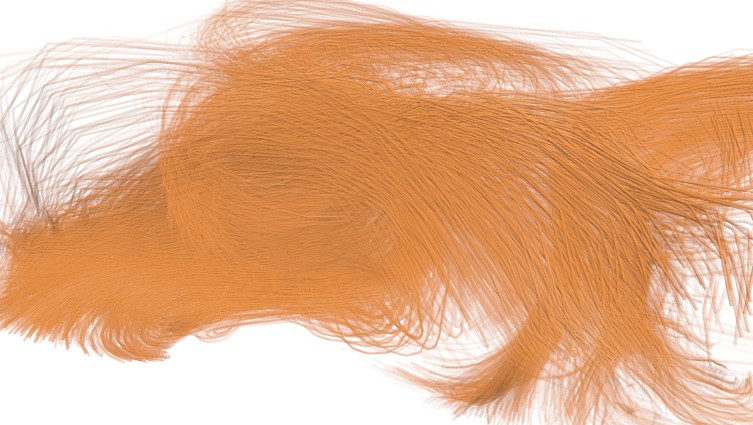}\hfill
\includegraphics[width=\VariableWidthIIII]{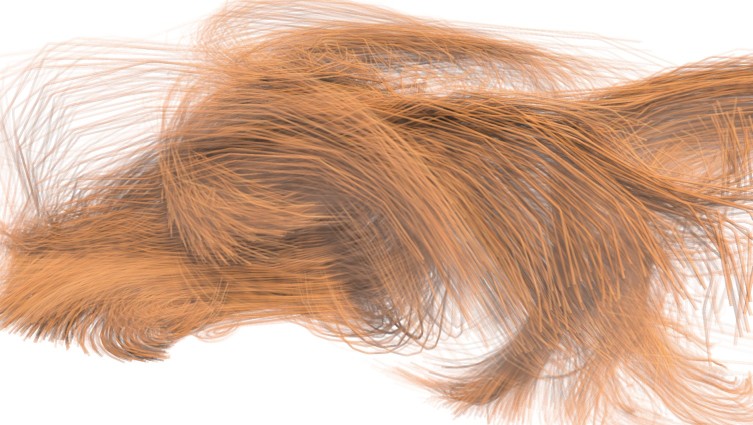}

\caption{First: Local illumination, opaque lines. Second: Global illumination, opaque lines. Third: Local illumination, semi-transparent lines, Fourth: Global illumination, semi-transparent lines. The Aneurysm II and Weather Forecast dataset are shown. 
}
\label{fig:colorpage-light}
\end{figure*}

\begin{figure*}

\includegraphics[width=\VariableWidthIII]{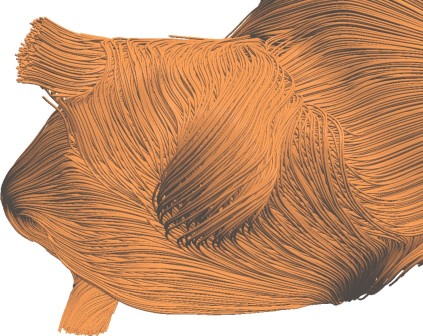}\hfill
\includegraphics[width=\VariableWidthIII]{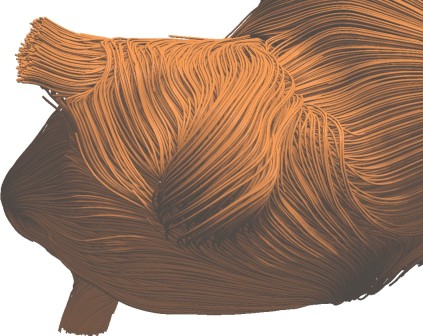}\hfill
\includegraphics[width=\VariableWidthIII]{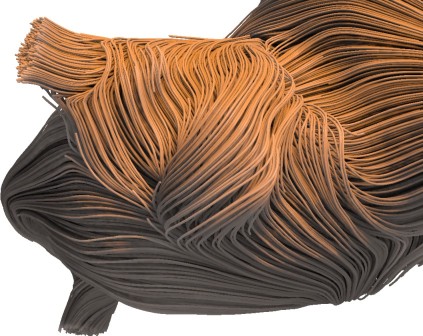}\hfill
 
\caption{First: Local illumination. Second: Soft shadows. Third: Soft shadows with ambient occlusion. The Aneurysm II dataset is shown. 
}
\label{fig:colorpage-light2}
\end{figure*}

\begin{figure*}

\includegraphics[width=\VariableWidthII]{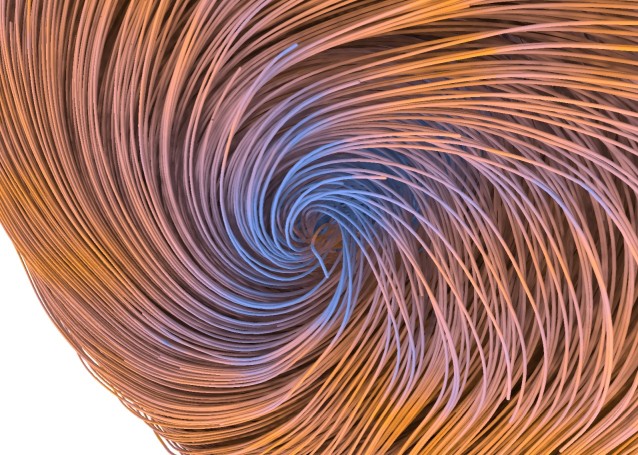}\hfill
\includegraphics[width=\VariableWidthII]{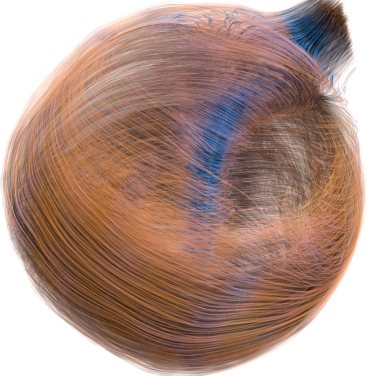}

\includegraphics[width=\VariableWidthII]{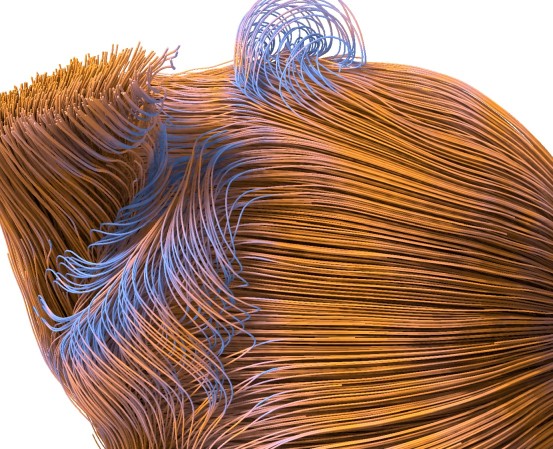}\hfill
\includegraphics[width=\VariableWidthII]{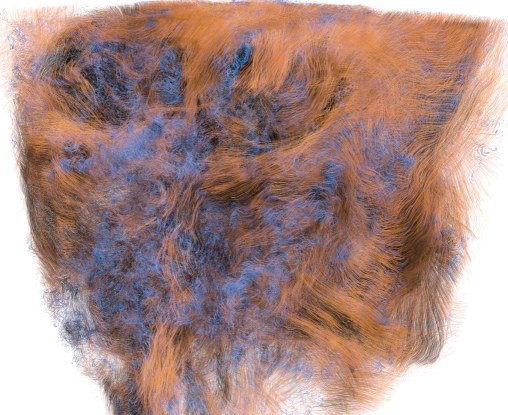}

\includegraphics[width=\VariableWidthII]{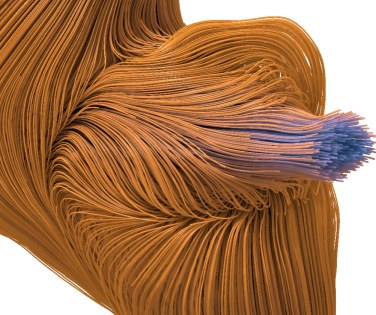}\hfill
\includegraphics[width=\VariableWidthII]{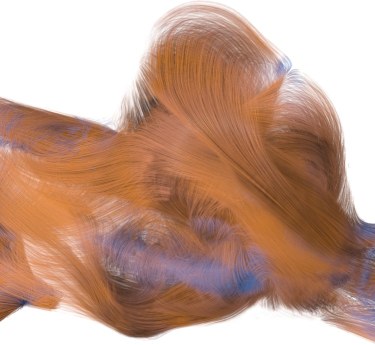}

\caption{Renderings of Aneurysm I (top and middle left), Turbulence (middle right), and Aneurysm II (bottom) }
\label{fig:colorpage3}
\end{figure*}